\documentclass[reprint, aps, prb, floatfix, superscriptaddress]{revtex4-2}

\pdfoutput=1
\usepackage{graphicx}
\usepackage{dcolumn}
\usepackage{bm}

\usepackage{amsmath,amssymb,amsfonts,verbatim, hyperref, mathtools, bbold, float}
\usepackage[utf8]{inputenc}

\usepackage[caption=false]{subfig}
\usepackage{physics}
\usepackage{soul, color}
\usepackage{algorithm, algorithmic}
\hypersetup{
	colorlinks=true,
	linkcolor=red,
	citecolor=blue,
	filecolor=blue,
}

\begin{document}

\title{Entanglement and thermalization in open fermion systems}

\author{Cristian \surname{Zanoci}}
\email{czanoci@stanford.edu}
\affiliation{Stanford Institute for Theoretical Physics, Stanford University, Stanford CA 94305}
\author{Brian \surname{Swingle}}
\email{bswingle@umd.edu}
\affiliation{Stanford Institute for Theoretical Physics, Stanford University, Stanford CA 94305}
\affiliation{Department of Physics, Harvard University, Cambridge MA 02138}
\affiliation{Martin Fisher School of Physics, Brandeis University, Waltham MA 02453}

\date{\today}

\begin{abstract}
We numerically study two non-interacting fermion models, a quantum wire model and a Chern insulator model, governed by open system Lindblad dynamics. The physical setup consists of a unitarily evolving ``bulk" coupled via its boundaries to two dissipative ``leads". The open system dynamics is chosen to drive the leads to thermal equilibrium, and by choosing different temperatures and chemical potentials for the two leads we may drive the bulk into a non-equilibrium current carrying steady state. We report two main results in this context. First, we show that for an appropriate choice of dynamics of the leads, the bulk state is also driven to thermal equilibrium even though the open system dynamics does not act directly on it. Second, we show that the steady state which emerges at late time, even in the presence of currents, is lightly entangled in the sense of having small mutual information and conditional mutual information for appropriate regions. We also report some results for the rate of approach to the steady state. These results have bearing on recent attempts to formulate a numerically tractable method to compute currents in strongly interacting models; specifically, they are relevant for the problem of designing simple leads that can drive a target system into thermal equilibrium at low temperature.
\end{abstract}

\maketitle


\section{Introduction}
\label{sec:intro}

One of the most challenging problems in quantum many-body physics is the calculation of electrical and thermal currents in strongly interacting systems~\cite{Datta1997,Blandin1976,Leggett1987,Han2013,Spohn1980}. The formalism of linear response provides a general method to compute currents in the limit of weak bias, but the needed correlation functions involve real time computations at finite temperature. When the coupling is strong these computations are typically intractable or rely on uncontrolled approximations. A variety of other methods can be brought to bear upon the problem, ranging from numerical approaches to the use of AdS/CFT duality~\cite{Schack1997,Tan1999,Vukics2007,Johansson2012,Vidal2004,Lothar2008,Hartnoll:2012rj, Mahajan:2013cja,Giombi:2011kc, GurAri:2012is,2008JHEP...02..045B}, but calculating transport properties of strongly interacting systems generally remains challenging. Given the wealth of experimental data on electrical and thermal currents in quantum many-body systems, it is important to address this ongoing challenge. We are particularly interested in low temperature dynamics, where collective modes can dominate the physics, e.g. at quantum critical points.

While a general approach to the problem of transport is difficult to imagine, recent ideas from the theory of quantum entanglement have offered some new hope in this direction. The starting point of this approach is to directly consider the physical properties of the state $\rho_{\text{NESS}}$ corresponding to the non-equilibrium steady state (NESS) of the system of interest (this is a starting point of many approaches, e.g.~\cite{Datta1997,Lothar2008,Pletyukhov2010,Egger2000,Blandin1976,Molmer1993,Dalibard1992,Gardiner1992,Prosen2008,Alicki2002,2016arXiv160806028G}). The steady state is imagined to carry a current at some finite temperature. Then while a general density matrix $\rho_{\text{NESS}}$ can be exponentially complex, it might be that the NESS has relatively little entanglement, similar to a thermal state~\cite{2008PhRvL.100g0502W,Verstraete2004}, and can be effectively compressed to a much smaller set of physically meaningful data. Working in the context of one-dimensional spin chains, it was indeed suggested that such NESS would have relatively little entanglement~\cite{Prosen2009}, e.g. that the mutual information between a region and its complement would obey an area law~\cite{Cubitt2015,doi:10.1063/1.4932612}. This insight led to an efficient procedure to compute currents in chaotic spin chains using matrix product state technology~\cite{Zwolak2004,Prosen2009,Blythe2007,White1992,Schollwock2005,Schollwock2011,Verstraete2004,Hartmann2009}.

There is evidence, from both numerics~\cite{Cui2015,Mascarenhas2015,Jin2013,Weimer2015,Cai2013,Kliesch2014,Prosen2010} and integrable systems~\cite{Karevski2013,Clark2010,Prosen2009,Prosen2012,Prosen2015,Prosen2011,Prosen2011-1,Prosen2014,Prosen2012-1,Popkov2015}, that matrix product states can give a good description of NESS in one dimension. However, we also know that not every NESS has low entanglement~\cite{PhysRevA.89.032321}. Recently, the framework of approximate conditional independence~\cite{petz1986,qmarkov,fr1,fr2,2015RSPSA.47150338W,2015arXiv150907127J} was used to argue that NESS of thermalizing systems would have an efficient tensor network representation in any dimension~\cite{mahajan2016entanglement}. The key idea is that in a thermalizing system, something like local thermal equilibrium is obtained in the NESS, at least if the system is driven only weakly away from equilibrium. If the entropic structure of such a local thermal equilibrium state follows reasonable expectations~\cite{2016arXiv160705753S}, then approximate conditional independence follows. Ref.~\cite{mahajan2016entanglement} gave a preliminary discussion of this physics within a free fermion open system model and using tools of AdS/CFT duality~\cite{1999IJTP...38.1113M,2006PhRvL..96r1602R,2008JHEP...02..045B}.

Although we are ultimately interested in physics of transport in interacting systems, the necessary algorithmic challenges have not yet been fully met. To help meet these challenges, non-interacting fermion models provide a useful testbed for some aspects of the physics. To set the stage, note that even if we grant that NESS have efficient tensor network representations, the non-trivial task of finding the right tensor network still remains. We cannot, for example, simply minimize the energy or the free energy of the system within a variational class of tensor network states since we are interested in an out-of-equilibrium state. One possible method for finding the steady state is a generalized variational principle adapted to out-of-equilibrium states~\cite{Cui2015}. Another method is to design an open system dynamics $\hat{\mathcal{L}}$ such that $\rho_{\text{NESS}}$ is equal to the time-independent steady state of $\hat{\mathcal{L}}$, $\hat{\mathcal{L}}(\rho_{\text{NESS}})=0$. The parameters defining the NESS, e.g. the temperature gradient, would be built into the dynamics $\hat{\mathcal{L}}$. One could then find the NESS by simulating the time evolution $\partial_t \rho = \hat{\mathcal{L}}(\rho)$ until it converges to the steady state. Of course, simulating the dynamics of $\hat{\mathcal{L}}$ may be challenging in general, but simulating the dynamics within a suitable class of low entanglement tensor network states could be feasible~\cite{Prosen2009,Zwolak2004,Yan2011,Stoudenmire2012,2016arXiv161200656K}.

Ref.~\cite{Prosen2009} found that a small bath consisting of just a few sites was sufficient to obtain interesting physics at high temperatures (compared to microscopic scales). It was also verified that the details of the bath did not strongly affect the results, again at high temperature. However, since we are particularly interested in low temperature physics, the problem of designing an open system dynamics whose steady state is the desired NESS is potentially non-trivial. To form a useful component of any computational method, such an open system dynamics must have three properties: (1) it must be able to drive the system to thermal equilibrium at low temperatures, (2) it must drive the system into the steady state in a reasonable (non-exponential) amount of time, and (3) it must not be excessively complex, e.g. it should not use the detailed properties of many-body energy eigenstates. In this work we investigate the problem of designing a suitable dynamics $\hat{\mathcal{L}}$ using a non-interacting fermion model.

The dynamical system has two components, illustrated in Fig.~\ref{fig:setup}: one is the system of interest (the ``target") and the other one is the designer leads which thermalize the target (or more generally drive it out of equilibrium). In this work, both the target and the lead are non-interacting fermion systems. We propose that when studying an interacting system, the target should include the interactions of the system of interest, but the lead can still be taken to be non-interacting. Based on the general expectation that interacting systems which thermalize do so regardless of the details of the bath, a non-interacting lead should be adequate to induce thermalization so long as it has the correct approximate features. Taking the lead to be non-interacting immediately answers point (3) above, since such a lead, as we review below, is relatively simple. In the remainder of this work, we study points (1) and (2) in both one- and two-dimensional non-interacting fermion models. The basic parameter of our lead model is the size of the lead; such a lead with many sites will be called an extended lead.

Our results are as follows. First, thermal equilibrium at relatively low temperatures can be reached using such an extended lead, but the required lead size typically grows with decreasing temperature. This addresses point (1) above. Second, the time to reach the steady state in a metallic state is an inverse polynomial in the system-plus-lead size. This addresses point (2) above. Third, the above two conclusions are modified when the system is insulating, either due to an energy gap or due to disorder. In these cases we find that the time to reach the steady state is significantly longer, scaling exponentially with the system size. Fourth, the steady states in our extended lead model are low entanglement states and exhibit approximate conditional independence for appropriate regions. To the best of our knowledge, we study lower temperatures and larger systems than previously considered in the literature. Compared to the results of Ref.~\cite{mahajan2016entanglement}, we study in more detail the structure of the steady state, both in the unbiased and biased case, and we study a two-dimensional model to investigate the validity of our conclusions in higher dimensions.

In the following sections we present our models, define our observables, and discuss our methods. We then present two sets of results, one for the one-dimensional quantum wire model and one for the two-dimensional Chern insulator model~\cite{qi2006topological}. We conclude with a brief discussion of future work and open questions.

\section{Models}
\label{sec:models}

\begin{figure}
	\includegraphics[width=\columnwidth]{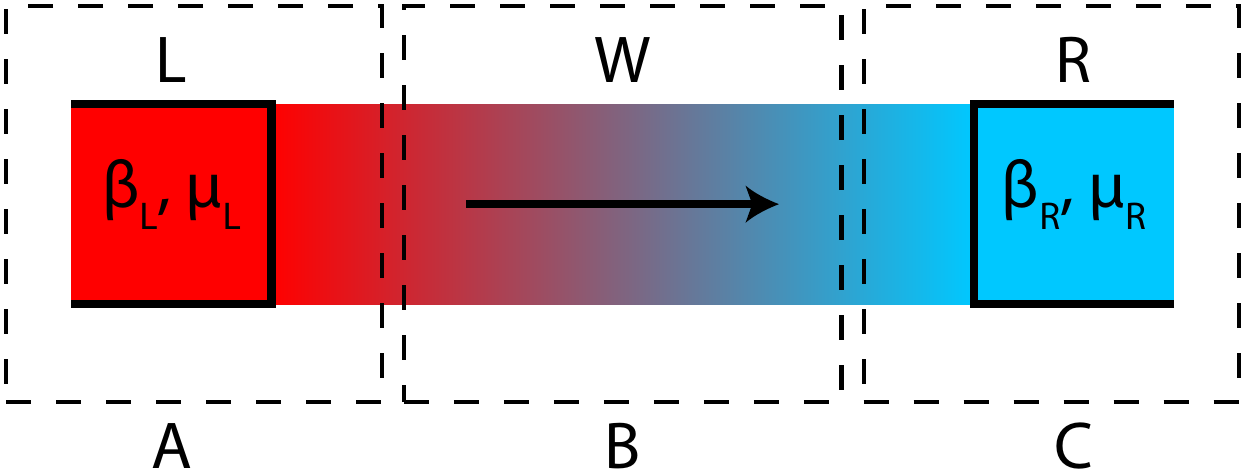}
	\caption{Schematic representation of the setup in which the total system is divided into two parts, the ``target" wire (W) in the middle and two ``leads" on the left (L) and right (R). The dashed boxes represent the partition of the system into three disjoint regions $A$, $B$, and $C$ for computing the mutual information $MI(A:C)$ and the conditional mutual information $CMI(A:C|B)$.}
	\label{fig:setup}
\end{figure}

We study the transport, entanglement, and thermalization properties of two models. The setup for each model consists of three disjoint regions: the left lead (L), the middle part representing the wire (W) whose transport properties we investigate, and the right lead (R) (see Fig.~\ref{fig:setup}). The wire and the leads are always arranged in this quasi-one-dimensional geometry. In the quantum wire model, the wire and leads are strictly one-dimensional, but in the Chern insulator model, the wire and leads are quasi-one-dimensional strips of a two-dimensional system. The three parts consist of $N_L$, $N_W$, and $N_R$ fermion sites respectively. The leads are held at inverse temperatures $\beta_L = 1/T_L$, $\beta_R = 1/T_R$ and chemical potential $\mu_L$, $\mu_R$. This is accomplished using open system dynamics which drives the decoupled leads to thermal equilibrium with the indicated parameters. We first describe the Hamiltonian dynamics and then discuss the implementation of dissipation.

\subsection{Hamiltonian dynamics}

\subsubsection{Quantum wire model}

The first model is a one-dimensional quantum wire of fermions. Each of the segments described above is characterized by a hopping Hamiltonian which is quadratic in fermion creation and annihilation operators

\begin{equation}
H_a = -w\sum_{x=1}^{N_a-1}(c_x^\dagger c_{x+1} + c_{x+1}^\dagger c_{x}),
\end{equation}
where $a = L, W, R$, and $c_x$ is the annihilation operator at site $x$. The creation and annihilation operators obey the standard anti-commutation relation $\{c_x, c_{y}^\dagger\} = \delta_{x, y}$. The leads couple to the middle part of the wire through similar hopping terms, but with a different coupling strength: $-w'(c_{N_L}^\dagger c_{N_L+1} + \text{h.c.})$ and $-w'(c_{N_L+N_W}^\dagger c_{N_L+N_W+1} + \text{h.c.})$. One can also add an on-site potential term $V_x c_x^\dagger c_x$ for every fermion in the middle region W.

\subsubsection{Chern insulator model}

The second model is a two-band tight-binding Hamiltonian in two dimensions which exhibits the physics of the quantum anomalous Hall effect. This model is called the Chern insulator model and was first studied in Ref.~\cite{qi2006topological}. We consider two states of fermions at each lattice site, which can be interpreted as $s$ and $p$ orbitals. The Hamiltonian in momentum space $\mathbf{k}=(k_x,k_y)$ is given by 

\begin{equation}
\label{eq:2D hamiltonian}
H_W = \sum_{\mathbf{k}} \left[ \epsilon(\mathbf{k}) + Vh(\mathbf{k}) \right],
\end{equation} 
where $\epsilon(\mathbf{k})$ and $h(\mathbf{k})$ are $2\times 2$ Hermitian matrices defined as

\begin{equation}
\epsilon(\mathbf{k})=
\begin{bmatrix}
-2t(\cos{k_x} + \cos{k_y}) & 0  \\
0 & -2t(\cos{k_x} + \cos{k_y})  \\
\end{bmatrix}
\end{equation}

\begin{equation}
h(\mathbf{k})=
\begin{bmatrix}
\tilde{h}(\mathbf{k}) & \sin{k_y} + i\sin{k_x}  \\
\sin{k_y}-i\sin{k_x} & -\tilde{h}(\mathbf{k})  \\
\end{bmatrix},
\end{equation}
with $\tilde{h}(\mathbf{k}) = c(2-\cos{k_x}-\cos{k_y}-e_s)$. We can perform a Fourier transform and write the Hamiltonian in position space

\begin{equation}
\begin{aligned}
H_W &= \sum_{x=1}^{N_{W, x}}\sum_{y=1}^{N_{W, y}} (c_{x, y}^\dagger H_1 c_{x, y+1} + \text{h.c.} \\
& + c_{x, y}^\dagger H_2 c_{x+1, y} + \text{h.c.} + c_{x, y}^\dagger H_3 c_{x, y}),
\end{aligned}
\end{equation}
and define each of the $2\times2$ matrices as

\begin{equation}
H_1 =\begin{bmatrix}
-\left(\dfrac{Vc}{2}+t\right) & \dfrac{iV}{2}  \\
\dfrac{iV}{2} & \left(\dfrac{Vc}{2}-t\right)  \\
\end{bmatrix},
\end{equation}

\begin{equation}
H_2 = \begin{bmatrix}
-\left(\dfrac{Vc}{2}+t\right) & -\dfrac{V}{2}  \\
\dfrac{V}{2} & \left(\dfrac{Vc}{2}-t\right)  \\
\end{bmatrix},
\end{equation}

\begin{equation}
H_3 =\begin{bmatrix}
Vc(2-e_s) & 0  \\
0 & -Vc(2-e_s)  \\
\end{bmatrix},
\end{equation}
where $N_{W, x}$, $N_{W, y}$ are the dimensions of the lattice and $c_{x, y} = [c_{x, y, 1}, c_{x, y, 2}]^T$ is a two-component column vector containing the annihilation operators of the two fermion states at lattice position $(x, y)$. We treat $\alpha = (x, y, i)$ as a composite index labeling the fermions, which obey the anti-commutation relation $\{c_{\alpha}, c_{\beta}^\dagger\} = \delta_{\alpha, \beta} = \delta_{x, x'}\delta_{y, y'}\delta_{i, i'}$. Note that the Hamiltonian contains only nearest-neighbor couplings and there are no periodic boundary conditions imposed.

An important feature of the above Hamiltonian is the existence of edge states. If we rewrite $H_W$ in the $(x, k_y)$ space, with periodic boundary conditions in y-direction and open boundary conditions in x-direction, we obtain a two-band energy spectrum with two edge states.

The lead Hamiltonians, $H_L$ and $H_R$, are also given by Eq.~\eqref{eq:2D hamiltonian}, where we set $V=0$ and $t=w$. The resulting lead Hamiltonian is similar to the $1D$ lead Hamiltonian because it involves only simple nearest neighbor hopping. The coupling between the left lead and the Chern insulator is given by

\begin{equation}
H_{LW} = -w'\sum_{y=1}^{N_{L, y}} c_{N_{L, x}, y}^\dagger J_2 c_{N_{L, x}+1, y} + \text{h.c.}
\end{equation}
where $J_2$ is a $2\times2$ all-ones matrix. An analogous term $H_{WR}$ can be written for the coupling between the Chern insulator and the right lead.

\subsection{Dissipative dynamics}

We describe the interaction of our leads with an environment using Markovian open system dynamics. The time evolution of the density matrix $\rho$ of the system is given by Lindblad's equation

\begin{equation}\label{mastereq}
\dfrac{d\rho}{dt} = \hat{\mathcal{L}}(\rho)\equiv  -i [H, \rho ] + \sum_{j} L_j \rho L_j^\dagger - \dfrac{1}{2}\sum_{j}\{L_j^\dagger L_j, \rho \},
\end{equation}
where $H$ is the full system Hamiltonian and $L_j$ are jump operators describing the coupling to the environment. The operator $\hat{\mathcal{L}}$ is called the Liouvillean super-operator and acts on the space of density matrix operators. Note that throughout this paper we set $\hbar = 1$.

We construct the jump operators such that, in the absence of coupling to the wire, each lead would be driven to the thermal equilibrium state appropriate for its decoupled lead Hamiltonian. The precise construction is discussed in Appendix B of Ref.~\cite{mahajan2016entanglement} and we review it here. The right and left jump operators are computed in the same way, so in what follows we focus only on the left lead. Recall that the single particle Hamiltonian of the left lead is quadratic and can be written as $H_L = c^\dagger h_Lc$, where $c = [c_1, \ldots, c_{N_L}]^T$. Let $\epsilon_j$ and $\psi_j$ be the eigenvalues and eigenvectors of $h_L$. The eigenvectors are written in the $\{c_\alpha\}_{\alpha = 1}^{N_L}$ basis and can be viewed as a function of lattice site. If we collect all the eigenvectors into a matrix $u_h = [\psi_1, \ldots, \psi_{N_L}]$ and all the eigenvalues into a diagonal matrix $d_h = \text{diag}(\epsilon_1, \ldots, \epsilon_{N_L})$, then we have $h_Lu_h = u_hd_h$. Therefore our Hamiltonian can be diagonalized as follows

\begin{equation}
H_L = c^\dagger u_hd_hu_h^\dagger c = (u_h^\dagger c)^\dagger d_h(u_h^\dagger c) = \overline{c}^\dagger d_h \overline{c} = \sum_{j = 1}^{N_L} \epsilon_j \overline{c}_j^\dagger  \overline{c}_j,
\label{eq:ham_diag}
\end{equation}
where we performed a change of basis $\overline{c}_j = \sum_{i = 1}^{N_L} \psi_{j, i}^* c_i$. Note that the new operators also satisfy the canonical anti-commutation relation $\{\overline{c}_x, \overline{c}_y^\dagger \}  = \psi_x^\dagger \psi_y = \delta_{x, y}$, due to the orthonormality of eigenvectors.

Now assume there are two Lindblad jump operators associated with each fermion mode $j$

\begin{equation}
L_{in, j} = \sqrt{\gamma_{in, j}} \cdot\overline{c}_j^\dagger,
\end{equation}

\begin{equation}
L_{out, j} = \sqrt{\gamma_{out, j}} \cdot\overline{c}_j,
\end{equation}
where $\gamma_{in, j}$ and $\gamma_{out, j}$ are the in and out rates. We want to fix these rates such that the thermal state $\rho_{th} = \exp(-\beta_L(H_L-\mu_L N_L))$ is a fixed point of the Lindblad equation~\eqref{mastereq}. The Hamiltonian term drops and we can decompose the master equation into $N_L$ independent equations of the form

\begin{equation}
\begin{aligned}
\gamma_{in, j}\overline{c}_j^\dagger & e^{-\beta_L(\epsilon_j - \mu_L)\overline{n}_j} \overline{c}_j +  \gamma_{out, j}\overline{c}_j e^{-\beta_L(\epsilon_j - \mu_L)\overline{n}_j} \overline{c}_j^\dagger  = \\ 
& \left( \gamma_{in, j}\overline{c}_j \overline{c}_j^\dagger + \gamma_{out, j} \overline{c}_j^\dagger \overline{c}_j \right) e^{-\beta_L(\epsilon_j - \mu_L)\overline{n}_j},
\end{aligned}
\end{equation}
with $\overline{n}_j = \overline{c}_j^\dagger\overline{c}_j$. As shown in Appendix B of Ref.~\cite{mahajan2016entanglement}, this equation simplifies to

\begin{equation}
\dfrac{\gamma_{in, j}}{\gamma_{out, j}} = e^{-\beta_L(\epsilon_j - \mu_L)}.
\end{equation}

Since only the ratio of the two rates matters, we can set $\gamma_{in, j} = \gamma$ and $\gamma_{out, j} = \gamma e^{\beta(\epsilon_j - \mu)}$ for all $j=1, 2, \ldots, N_L$. Substituting this back into the initial equations for the jump operators, we arrive at the final form of our Lindblad operators

\begin{equation}
L_{in, j} = \sqrt{\gamma} \sum_{i=1}^{N_L}\psi_{j, i}c_i^\dagger,
\end{equation} 

\begin{equation}
L_{out, j} = \sqrt{\gamma e^{\beta_L(\epsilon_j - \mu_L)}} \sum_{i=1}^{N_L}\psi_{j, i}^*c_i.
\end{equation} 
Similar formulas are also obtained for the right lead. 

The most important features of these jump operators are that they are linear in the fermionic creation and annihilation operators and that they thermalize the corresponding leads in the absence of contact with the wire. It is worth mentioning that in our case of a non-interacting fermion system with jump operators linear in the fermion modes, the Lindblad equation~\eqref{mastereq} reduces to a single particle equation for the Green's function~\cite{2013NJPh...15h5001B}, as reviewed in Appendix A of Ref.~\cite{mahajan2016entanglement}.

\subsection{Observables}
\label{sec:obs}

For each of the two models defined above, we compute several observables that reveal the transport and entanglement properties of non-equilibrium steady states. We study electrical currents to probe out-of-equilibrium physics, energy occupation numbers to probe thermalization, decay rates of the open system dynamics to probe convergence timescales, and mutual information and conditional mutual information to probe the entanglement structure of the state.

\subsubsection{Currents}

For the one-dimensional wire, the current operator through the link $(j, j+1)$ is 

\begin{equation}
I = -iw(c_j^\dagger c_{j+1} - c_{j+1}^\dagger c_j).
\end{equation} 
For the two-dimensional lattice, we define the current flowing in the $x$-direction across the link $((x, y), (x+1, y))$

\begin{equation}
I_x = ic_{x, y}^\dagger H_2 c_{x+1, y} + \text{h.c.}
\end{equation} 
and the current flowing in the $y$-direction across the link $((x, y), (x, y+1))$

\begin{equation}
I_y = ic_{x, y}^\dagger H_1 c_{x, y+1} + \text{h.c.}
\end{equation} 

\subsubsection{Occupation numbers} 

Next we are interested in computing the occupation numbers of the energy eigenstates of the bulk in NESS and comparing them to the thermal equilibrium distribution. We begin by writing the bulk Hamiltonian $H_{W}$ as a $N_W\times N_W$ matrix in the $\{c_j\}$ basis. Let $\epsilon_k$ and $\psi_k$ be the eigenvalues (single-particle energies) and eigenvectors of $H_W$. Then, for each energy mode $k$, we can define an annihilation operator 

\begin{equation}
c_{\epsilon_k} = \sum_{j=1}^{N_W} \psi_k^*(j)c_j ,
\end{equation} 
where the sum is over all the fermion modes in the bulk. The number operator for energy mode $k$ is given by

\begin{equation}
\langle c_{\epsilon_k}^\dagger c_{\epsilon_k} \rangle= \sum_{i=1}^{N_W}\sum_{j=1}^{N_W} \psi_k(i)\psi_k^*(j) \langle c_i^\dagger c_j \rangle.
\end{equation} 
We will compare these expectation values in the exact thermal state with those in the steady state of the open system.

\subsubsection{Decay rates} 

We will also compute the rate of relaxation $\Delta$ to the steady state. This is obtained from the spectrum of the Liouvillean operator $\hat{\mathcal{L}}$ as discussed in Sec.~\ref{sec:methods} and Appendix~\ref{sec:Appendix A}. The inverse of this rate determines the time needed to come exponentially close to the steady state. The rate typically decreases with increasing system size $n$, either as an inverse polynomial $\Delta \sim 1/n^a$ or exponentially $\Delta \sim e^{- b n}$.

\subsubsection{Entanglement} 

Finally, to study the entanglement structure of the current-carrying states we compute the mutual information and conditional mutual information. If we consider two regions, $A$ and $B$, of our system, then the mutual information is defined as

\begin{equation}
MI(A : B) = S(A) + S(B) - S(AB) ,
\end{equation} 
where $S(X)$ denotes the von Neumann entropy of region $X$. The conditional mutual information between three regions, A, B, and C (see Fig.~\ref{fig:setup}), of a system is given by

\begin{equation}
CMI(A : C | B) = S(AB) + S(BC) - S(B) - S(ABC).
\end{equation} 
Notice that the quantities defined above are expressed in terms of von Neumann entropy only. To compute the entropy of a region $X$, we first define the correlation matrix 

\begin{equation}
G^X_{\alpha \beta} = \langle c_\alpha^\dagger c_\beta \rangle ,
\end{equation}
where $\alpha$ and $\beta$ denote orbitals within $X$. It is worth mentioning that $G^X$ has eigenvalues between $0$ and $1$. We can then define the entropy as

\begin{equation}
S(X) = -\text{Tr}(G^X\ln{G^X} + (1-G^X)\ln(1-G^X)) .
\end{equation}

The interpretation of vanishing mutual information is simple. If $MI(A:C) =0$ then $\rho_{AC} = \rho_A \otimes \rho_C$ and the subsystems $A$ and $C$ are uncorrelated. The interpretation of vanishing conditional mutual information is more subtle. If $CMI(A:C|B)=0$ then $A \rightarrow B \rightarrow C$ forms a ``quantum Markov chain"~\cite{qmarkov} meaning that $C$ is independent of $A$ given the state of $B$. As discussed at length in Refs.~\cite{petz1986,qmarkov,fr1,fr2,2015RSPSA.47150338W,2015arXiv150907127J,2016arXiv160705753S,mahajan2016entanglement}, a quantum Markov chain generalizes the classical notion of a Markov chain in the sense that the total state $\rho_{ABC}$ can be recovered from the marginals $\rho_{AB}$ and $\rho_{BC}$. This does not imply that $A$ is uncorrelated with $C$, but it does imply that they are unentangled. Hence vanishing conditional mutual information indicates a certain kind of short-range entanglement in the state. This concept of approximate conditional independence was used in Refs.~\cite{2016arXiv160705753S,mahajan2016entanglement} to show that certain NESS have tensor network representations.

\section{Methods}
\label{sec:methods}

In order to compute our observables, we use a technique developed in Ref.~\cite{prosen2008third}, which allows us to solve the Lindblad master equation using a canonical quantization in the Fock space of operators. The method is applicable under the condition that the Hamiltonian is quadratic and the jump operators $L_j$ are linear in fermionic operators. The key idea is to write the Liouvillean $\hat{\mathcal{L}}$ in terms of adjoint Majorana maps and diagonalize it in the basis of normal master modes, which represent anti-commuting super-operators acting on the Fock space of density operators. The physical NESS is given by the zero-mode eigenvector. Below we review the main results of Ref.~\cite{prosen2008third}. A more detailed discussion of the method is included in Appendix \ref{sec:Appendix A}.

We begin by rewriting the canonical fermion operators in terms of Majorana operators

\begin{align}
w_{2j-1} = c_j+c_j^\dagger, && w_{2j} = i(c_j - c_j^\dagger),
\end{align}	
satisfying the anti-commutation relation $\{w_j, w_k\} = 2\delta_{j, k}$. Throughout this section the labels $i,j,...$ run over all fermion modes in the problem, spatial and otherwise. Then the Hamiltonians of our systems can be written as quadratic forms in Majorana fermions

\begin{equation}
\label{eq:Hamiltonian}
H = \sum_{j, k = 1}^{2n}w_jH_{jk}w_k,
\end{equation}	
while the Lindblad operators can be written as linear combinations of $w_j$

\begin{equation}
\label{eq:bath}
L_j = \sum_{k = 1}^{2n}l_{j, k}w_k,
\end{equation}
where $n = N_L + N_W + N_R$ is the total system size.

Ref.~\cite{prosen2008third} shows that the properties of the NESS can be derived from the Liouvillean shape matrix $A$, which is an antisymmetric $4n\times 4n$ matrix incorporating the parameters of the Hamiltonian and Lindblad operators

\begin{equation}
\label{eq:matrix A}
\begin{matrix*}[l]
A_{2j-1, 2k-1} &=&-2iH_{jk} -M_{kj} + M_{jk} \\
A_{2j-1, 2k} &=&\hphantom{-}2iM_{jk} \\
A_{2j, 2k-1} &=&-2iM_{kj} \\
A_{2j, 2k} &=&-2iH_{jk} -M_{jk} + M_{kj}
\end{matrix*}
\end{equation}
where $M$ is a Hermitian matrix with entries

\begin{equation}
\label{eq:matrix M}
M_{jk} = \dfrac{1}{2}\sum_{i}l_{i, j}l_{i, k}^* .
\end{equation}
It is worth mentioning that our formula for $A$ is different from the one in Ref.~\cite{prosen2008third} in that we swap the indices of $M$~\footnote{There is also an extra factor of $\frac{1}{2}$ in the definition of $M$ which comes from a rescaling of Lindblad operators $L_k\rightarrow \frac{L_k}{\sqrt{2}}$ relative to the definition in Ref.~\cite{prosen2008third}.}. We believe this merely reflects a typo in Ref.~\cite{prosen2008third}. 

Recall that the eigenvalues of a complex antisymmetric matrix of even dimension always come in pairs $\pm \beta$. Therefore we can order the eigenvalues of $A$ as $\beta_1, -\beta_1, \beta_2, -\beta_2, \ldots, \beta_{2n}, -\beta_{2n}$, with $\operatorname{Re}(\beta_1)\geq \operatorname{Re}(\beta_2)\geq\ldots\geq \operatorname{Re}(\beta_{2n}) \geq 0$. Let $v_1, v_2, \ldots, v_{4n}$ be the corresponding eigenvectors, written as column vectors.

Refs.~\cite{prosen2008third, prosen2010exact} prove three key results that we use to compute our observables. If the eigenvalues of $A$ have strictly positive real parts, $\operatorname{Re}(\beta_j) > 0$, then 

\begin{enumerate}
	\item The non-equilibrium steady state is unique.
	\item The rate of exponential relaxation to NESS is given by $\Delta = 2\operatorname{Re}(\beta_{2n})$.
	\item The expectation value of any quadratic observable $w_jw_k$ in NESS is given by
	
	\begin{equation}
	\langle w_jw_k \rangle = 2\sum_{m=1}^{2n} v_{2m, 2j-1}v_{2m-1, 2k-1} .
	\end{equation}
\end{enumerate}

These results may be understood intuitively by noting that, roughly speaking, the $\beta_j$ represent the ``energies of excitations". Hence the NESS is unique when all $\beta_j$ have non-zero real part because all other states decay. Similarly, the rate of relaxation is determined by the slowest decaying excitation, corresponding to $\beta_{2n}$ in the ordering we have chosen.

Notice that the results above provide all the necessary information to compute our observables, since they are all expressed in terms of the two-point correlation functions $\langle c_j^\dagger c_k \rangle$, which of course is quadratic in Majorana operators. One can further use Wick's theorem to compute expectation values of any higher order observable with an even number of fermion operators.

The implementation of these techniques involves computing the eigenvalue decomposition of very large matrices. We used the Multiprecision Computing Toolbox for MATLAB to solve for the eigenstates and eigenvalues of $A$ with high precision ($32$ digits). The extra precision was required to ensure that we can correctly group the eigenvalues into $(\beta, -\beta)$ pairs and that the eigenvectors are orthonormal.

\section{Results}
\label{sec:res}

\subsection{Quantum wire model}
\label{sec:wire}

In this subsection we describe our analysis of the steady state physics of the quantum wire model. Throughout this discussion we set the nearest neighbor Hamiltonian couplings to be $w=w'=1$, so that we effectively measure all energies in units of $w$ and all times in units of $1/w$. We first study the physics of thermalization when the leads are unbiased. Then we consider the physics of NESS when the leads are biased. Next we discuss the structure of entanglement in the steady state. Finally, we discuss the effects of disorder.

\subsubsection{Unbiased thermal equilibrium}

We begin our analysis of the quantum wire model by investigating thermalization with unbiased leads. Therefore we set $\beta_L = \beta_R = \beta$ and $\mu_L = \mu_R = 0$. Recall that the open system dynamics is designed to drive each decoupled lead (L and R) to its decoupled thermal state. However, the leads are coupled to the middle wire which, except for the lead coupling at its boundaries, enjoys unitary dynamics. We try to assess how well the wire is thermalized by its boundary couplings to the leads. 

\begin{figure}
    \centering
    \subfloat[]{\includegraphics[width=\columnwidth]{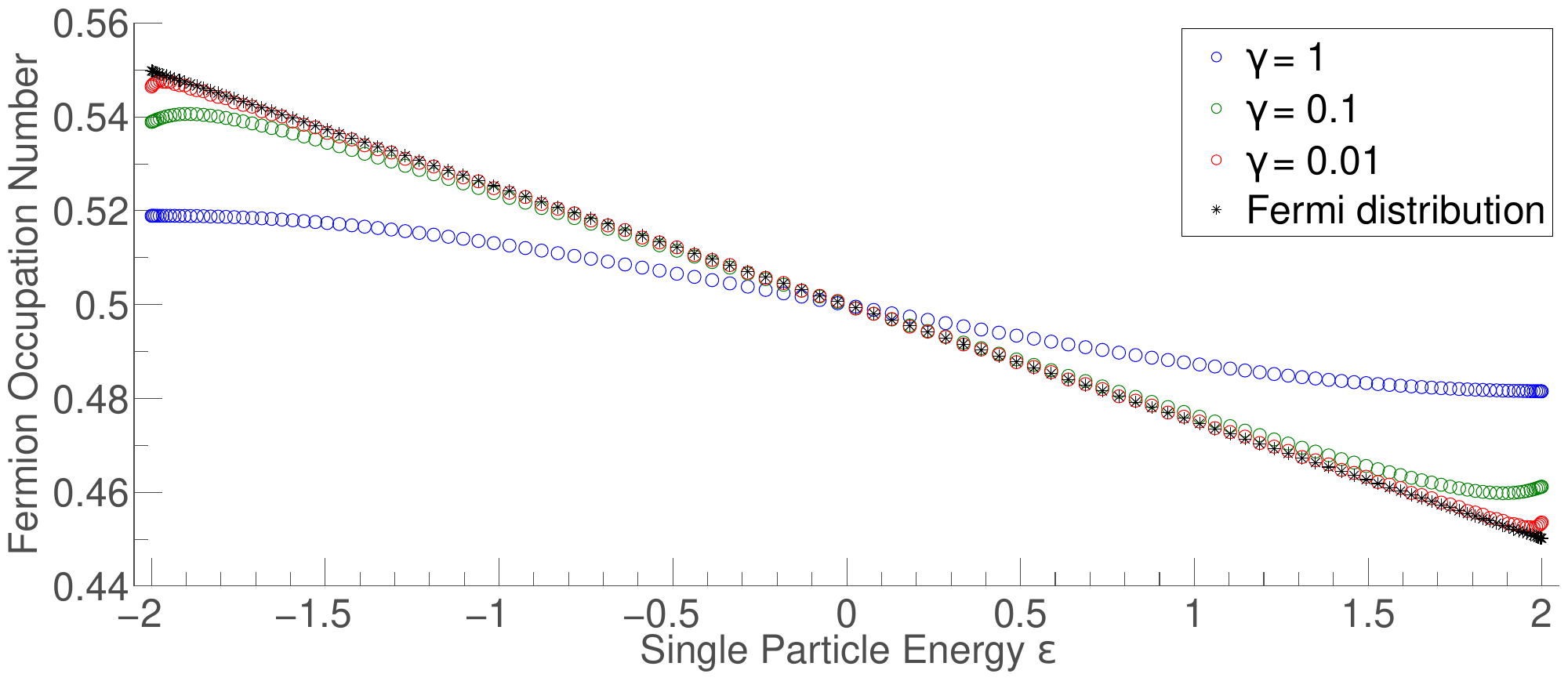}}
    
    \subfloat[]{\includegraphics[width=\columnwidth]{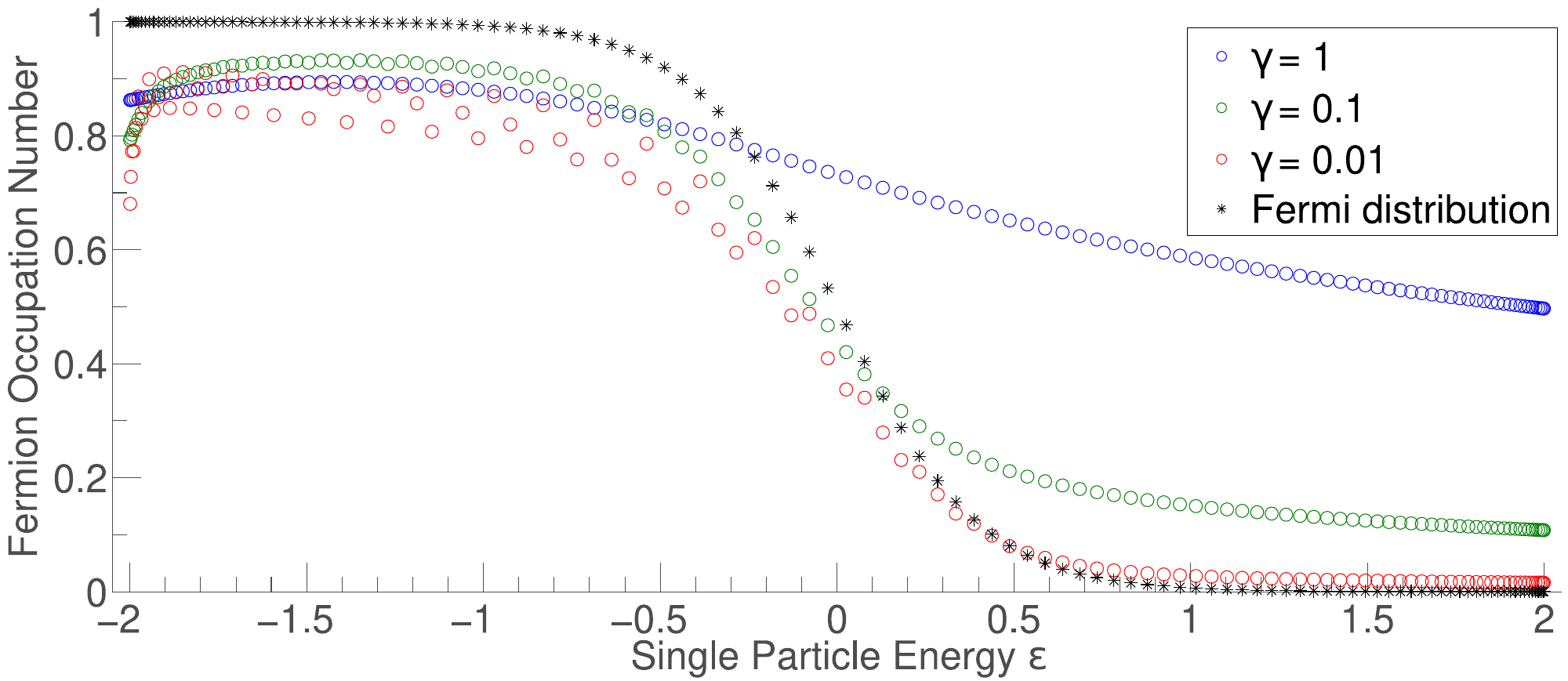}}
	\caption{Comparison of energy eigenstate occupation numbers between thermal equilibrium and the steady state of $\hat{\mathcal{L}}$. Occupation numbers for (a) $\beta = 0.1$ and (b) $\beta = 5$. System sizes are $N_L=N_R=40$ and $N_W=120$.}
	\label{fig:occupation_numbers_1D}
\end{figure}

We compare the exact thermal occupation numbers of the decoupled wire ($w'=0$) with the expectation values of the corresponding number operators in the steady state ($w'=1$). We expect agreement if (1) the leads are effectively thermalizing the wire and (2) the effects of the lead-wire coupling $w'$ is small. To be clear, throughout the analysis we use the decoupled ($w'=0$) wire energy eigenstates and we compare the expectation value of the number operator in thermal equilibrium and in the steady state of the open system dynamics. Data for $\beta = 0.1$ and $\beta = 5$ as a function of $\gamma$ (the strength of the dissipative terms in $\hat{\mathcal{L}}$) is shown in Fig.~\ref{fig:occupation_numbers_1D}. 

\begin{figure} 
    \centering
    \subfloat[]{\includegraphics[width=\columnwidth]{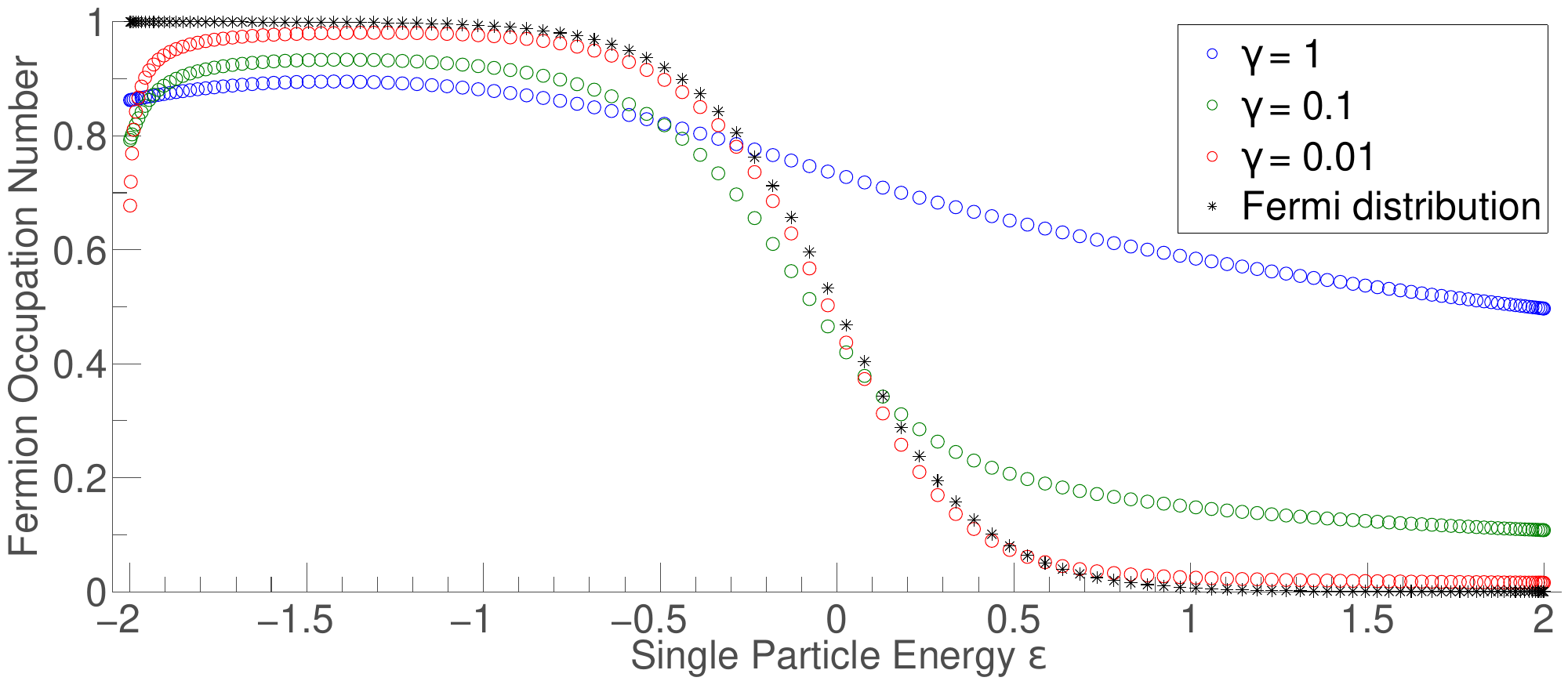}}
    
    \subfloat[]{\includegraphics[width=\columnwidth]{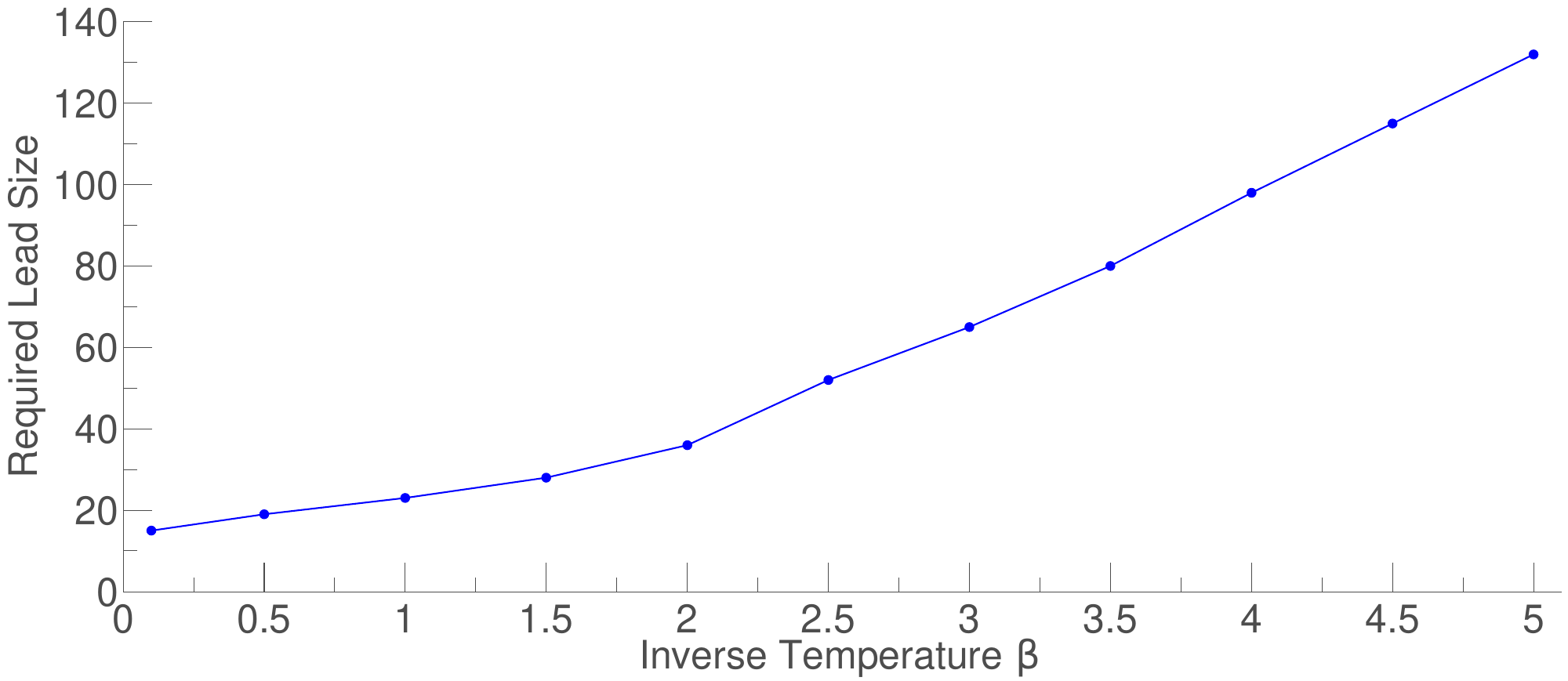}}
	
	\caption{(a) Energy eigenstate occupation numbers for $\beta = 5$ and $N_L=N_R=N_W=120$. (b) The minimal size $N_L = N_R$ of the leads required to approximately thermalize the wire ($N_W = 120$) as a function of $\beta$. Recall that we do not expect exact agreement with the decoupled ($w'=0$) thermal result anyway since $w'\neq 0$ in the Liouvillean $\hat{\mathcal{L}}$.}
	\label{fig:lead_size_vs_beta}
\end{figure}

For the high-temperature regime $\beta =0.1$, we find that $\gamma = 0.01$ leads to a steady state distribution of occupation numbers that agrees well with the thermal result. Increasing the lead size does not dramatically affect the final steady state. For the lowest temperatures reached, $\beta = 5$, a lead of size $N_L = N_R = 40$ is not sufficient to thermalize a wire of size $N_W=120$. As demonstrated in Fig.~\ref{fig:lead_size_vs_beta}(a), by taking larger leads we are able to achieve thermal equilibrium even for temperatures as low as $\beta = 5$.

Next, we investigate the lead size required to have approximate thermalization of the wire at a given inverse temperature $\beta$. Figure \ref{fig:lead_size_vs_beta}(b) shows the minimal size $N_L=N_R$ of the leads needed in order for the occupation numbers to be close (within a few percent) to the thermal distribution for different temperatures. Note again that the size of the contacts is only important at low temperatures. Similar results are also obtained when studying energy eigenstates of the entire coupled wire and lead system. We conclude that a sufficiently large lead is able to thermalize a wire even at low temperature only via boundary couplings.

\subsubsection{Biased steady state}

We continue our analysis by probing the extent to which biased leads define a steady state with the expected transport physics. To this end we take $\beta_L = \beta_R = \beta$ and $\mu_L = - \mu_R$. We compare the value of the current flowing through the wire in steady state with the corresponding Landauer formula for conductance~\cite{5392683}. Recall that this formula is obtained from an infinite wire model where the left and right moving particles are emitted from separate reservoirs with potentially different temperatures and chemical potentials.

The linear response conductance is defined as $G = I / (\mu_L - \mu_R)$ for small $\mu_L - \mu_R$. We study the conductance as a function of inverse temperature $\beta$ by computing the current in the middle of the chain in the steady state with weakly biased leads. Note that the current is uniform throughout the chain in a stationary state. As reviewed in Appendix D of Ref.~\cite{mahajan2016entanglement}, the current predicted by the ballistic Landauer formula is

\begin{equation}
    I = \int_0^\pi \dfrac{dk}{2\pi} v_k f(\epsilon_k-\mu_L, \beta_L) + \int_{-\pi}^0 \dfrac{dk}{2\pi} v_k f(\epsilon_k-\mu_R, \beta_R),
\end{equation}
where $\epsilon_k = -2w\cos{k}$ is the energy, $v_k = \frac{\partial \epsilon_k}{\partial k} = 2w\sin{k}$ is the group velocity, and $f(\epsilon, \beta) = (e^{\beta\epsilon}+1)^{-1}$ is the Fermi distribution. This can be explicitly integrated to give 

\begin{equation}
\begin{aligned}
    I = \dfrac{1}{2\pi} \bigg(&\dfrac{1}{\beta_L}\ln{\left( \dfrac{e^{\beta_L(\mu_L+2w)}+1}{e^{\beta_L(\mu_L-2w)}+1}\right)} + \\ & \dfrac{1}{\beta_R}\ln{\left( \dfrac{e^{\beta_R(\mu_R-2w)}+1}{e^{\beta_R(\mu_R+2w)}+1}\right)}\bigg).
\end{aligned}
\end{equation}

\begin{figure} 
	\centering
	\includegraphics[width=\columnwidth]{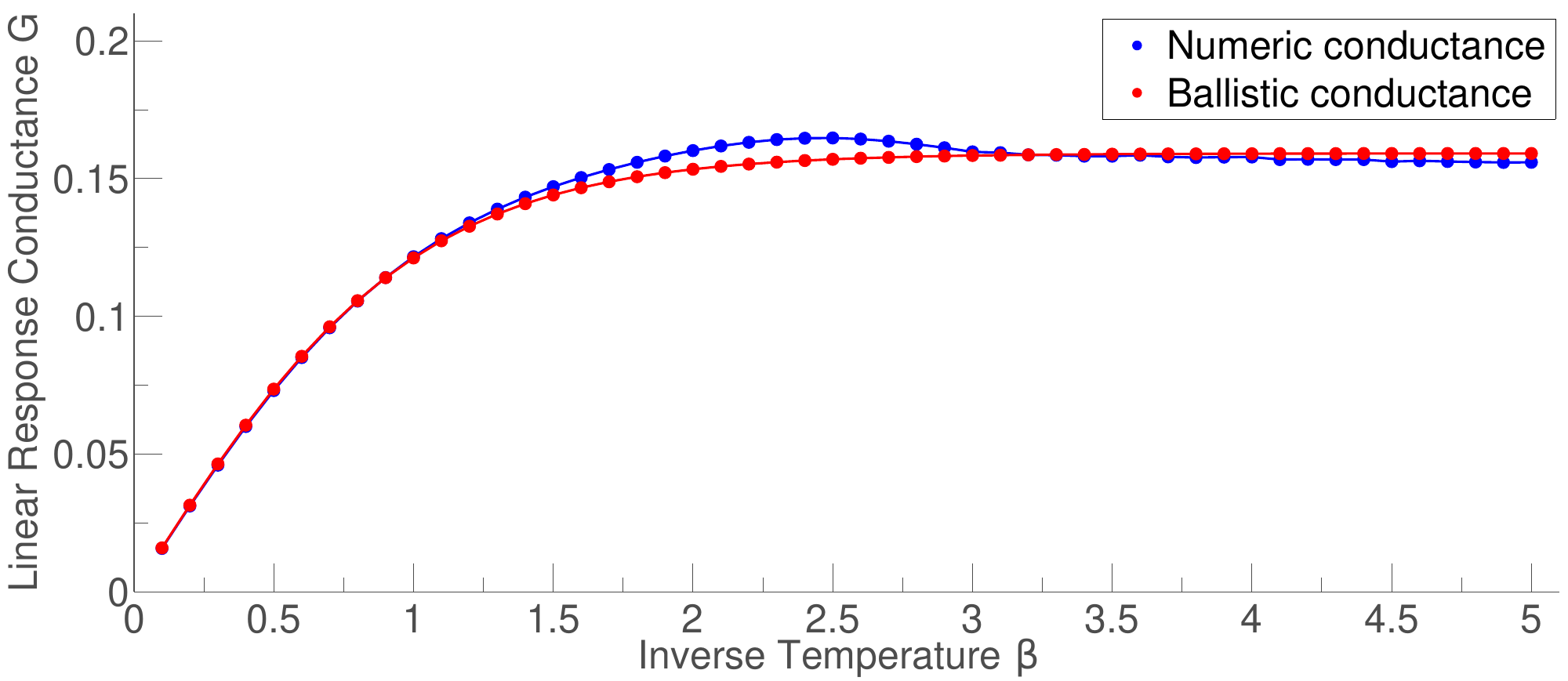}
	\caption{Landauer formula for ballistic conductance (red) compared to the numerically evaluated conductance in the steady state of $\hat{\mathcal{L}}$ (blue) as a function of $\beta$. System parameters are $N_L=N_W=N_R=80$, $w=w'=1$, and $\gamma = 0.05$.}
	\label{fig:conductance}
\end{figure}

Next, in Fig.~\ref{fig:conductance} we plot the ballistic transport conductance and the numerically computed conductance in the steady state as a function of $\beta = \beta_L=\beta_R$, for a system with $240$ sites ($N_L=N_W=N_R=80$) and $\gamma = 0.05$. Although there is an interesting feature near $\beta=2$, the steady state current generally matches the Landauer formula quite well. We also obtained numerical evidence that the low temperature Landauer result is well approximated by the steady state value in the limit of large leads. Together we take this data as evidence that, just as the unbiased steady state captures well the physics of thermal equilibrium, the biased steady state captures well the physics of current carrying states in the non-interacting fermion wire.

\begin{figure}
    \centering
    \subfloat[]{\includegraphics[width=\columnwidth]{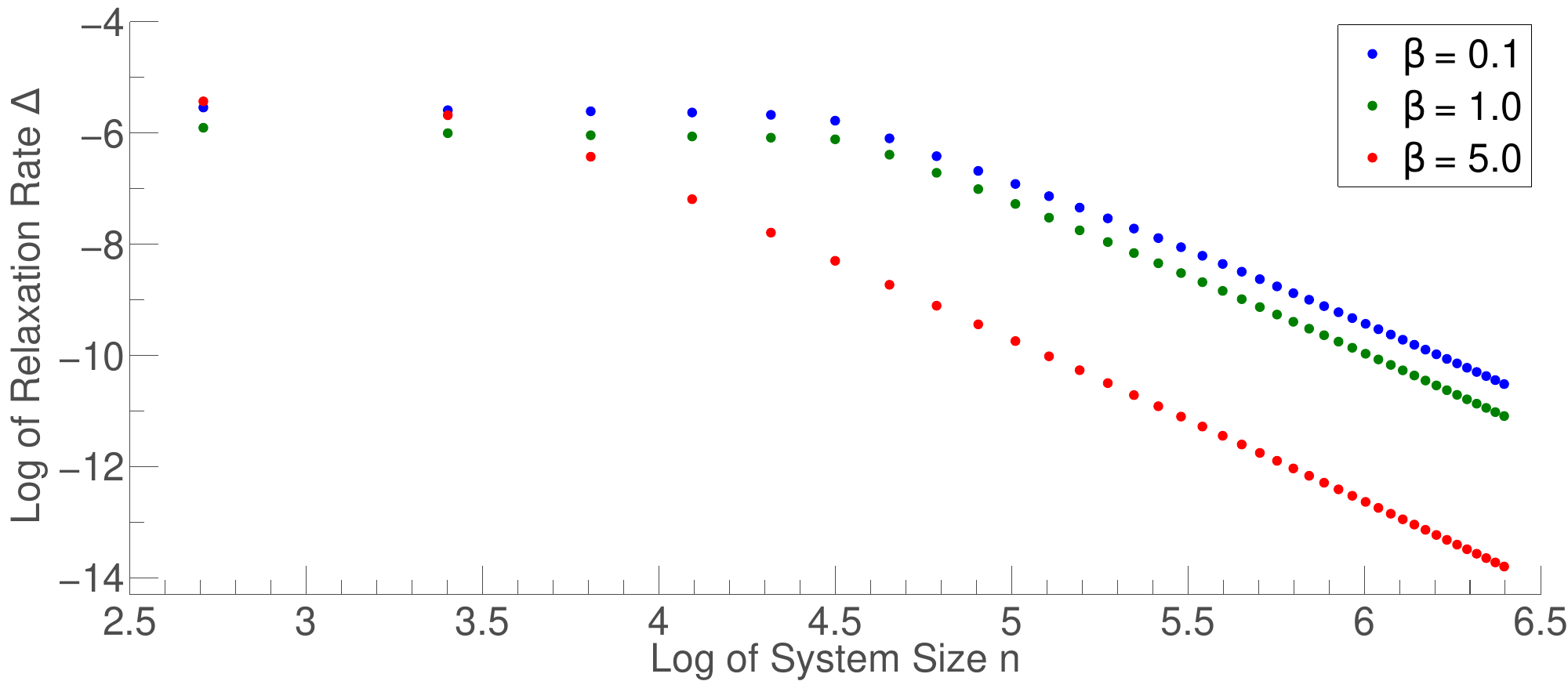}}
    
    \subfloat[]{\includegraphics[width=\columnwidth]{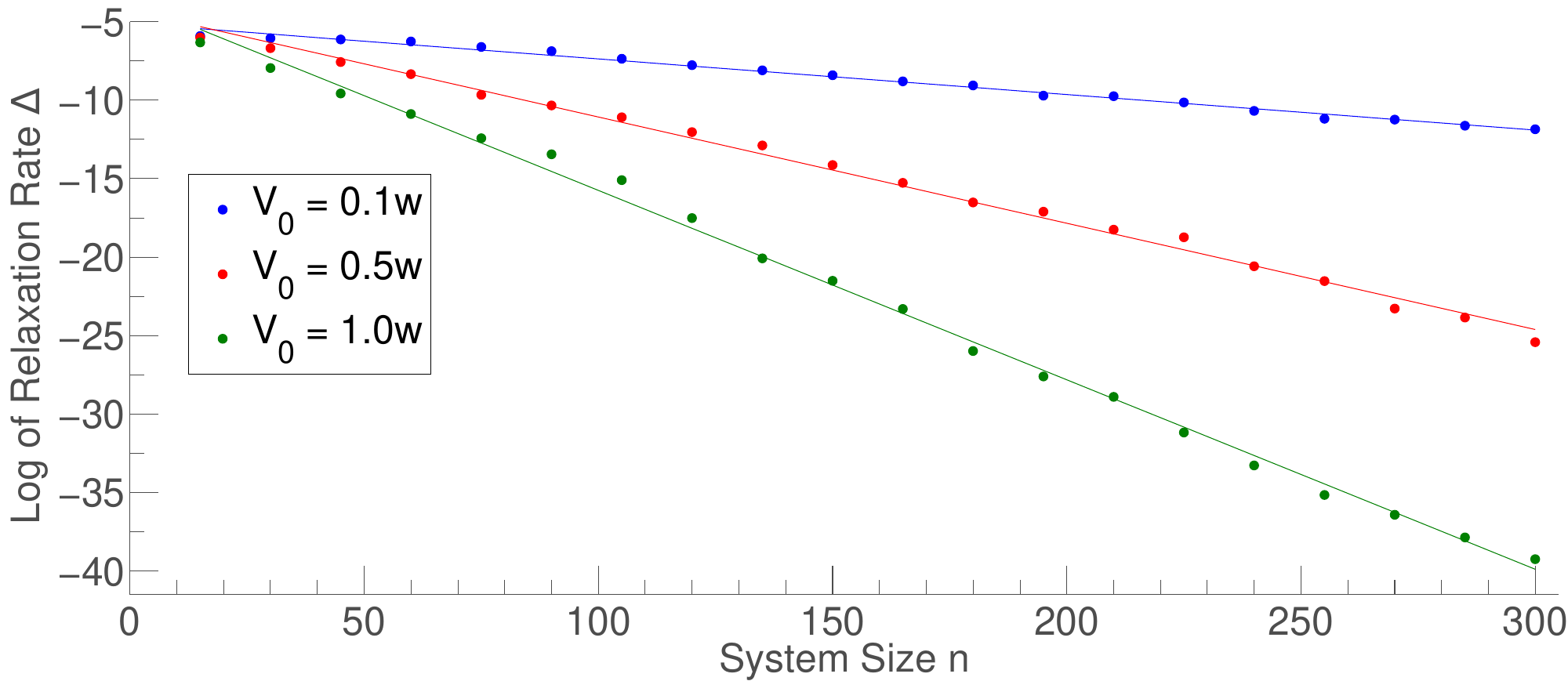}}
	
	\caption{ Log of the relaxation rates $\Delta$ as a function of system size $n=N_L + N_W + N_R$ for (a) the translation invariant system with $V_0 = 0$ and (b) the disordered system with different disorder strengths $V_0$. Averaging is performed over $1000$ disorder realizations. Note that unlike in the translation invariant case, where $\Delta \sim n^{-3}$, in the disordered case the decay rate $\Delta$ scales exponentially $\Delta \sim e^{-b n}$ with the total system size $n$.}
	\label{fig:relaxation_rates}
\end{figure}

We can also show that relaxation to the steady state takes place in a reasonable time frame, inverse polynomial in the system size. Recall that the relaxation rate is determined from the real part of the spectrum of the Liouvillean as described in Sec.~\ref{sec:methods}. Figure~\ref{fig:relaxation_rates}(a) shows the relaxation rate $\Delta$ as a function of total system size $n$ for different inverse temperatures $\beta$. Other parameters are $\gamma = 0.05$ and $\mu_R = -\mu_L = 0.1$. We clearly observe an inverse polynomial scaling at large system sizes. 

\subsubsection{Entanglement structure of the steady state}

\begin{figure}
    \centering
    \subfloat[]{\includegraphics[width=\columnwidth]{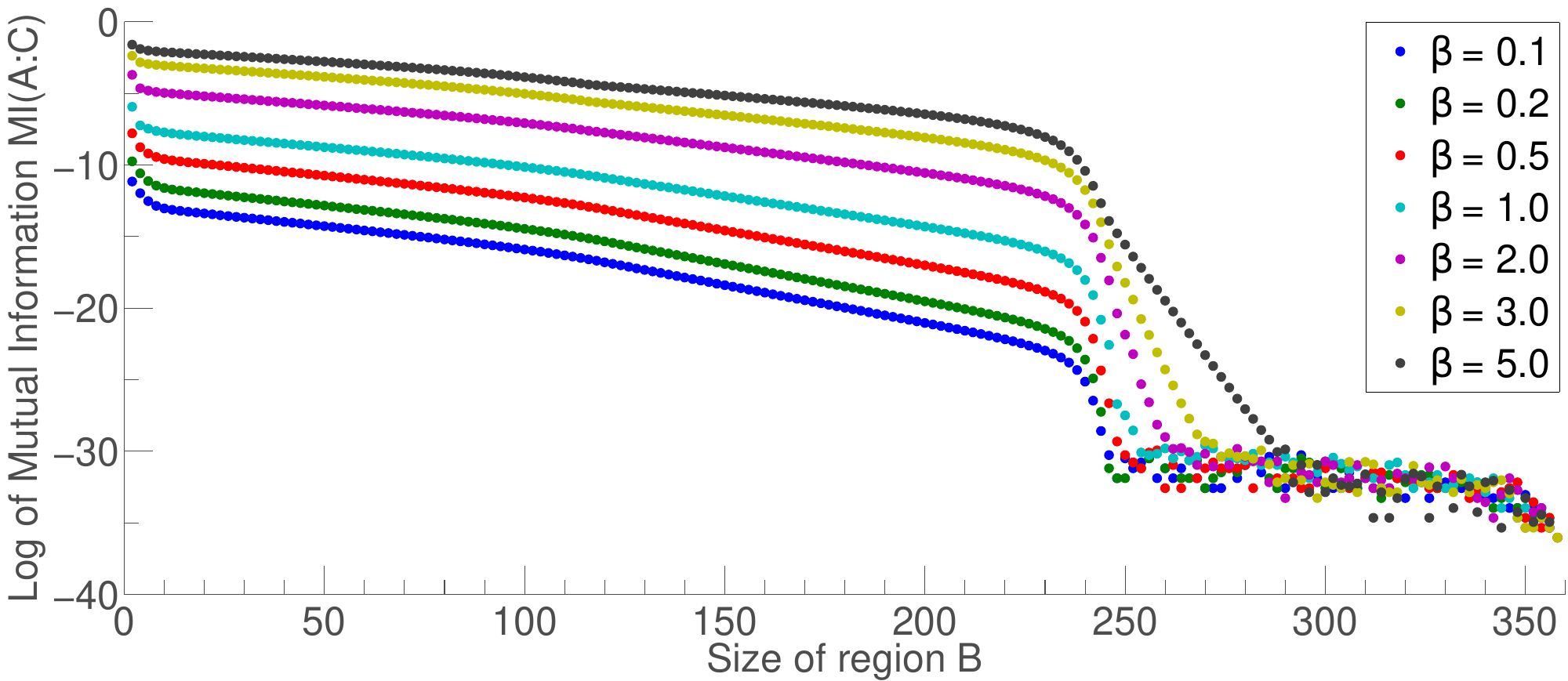}}
	
	\subfloat[]{\includegraphics[width=\columnwidth]{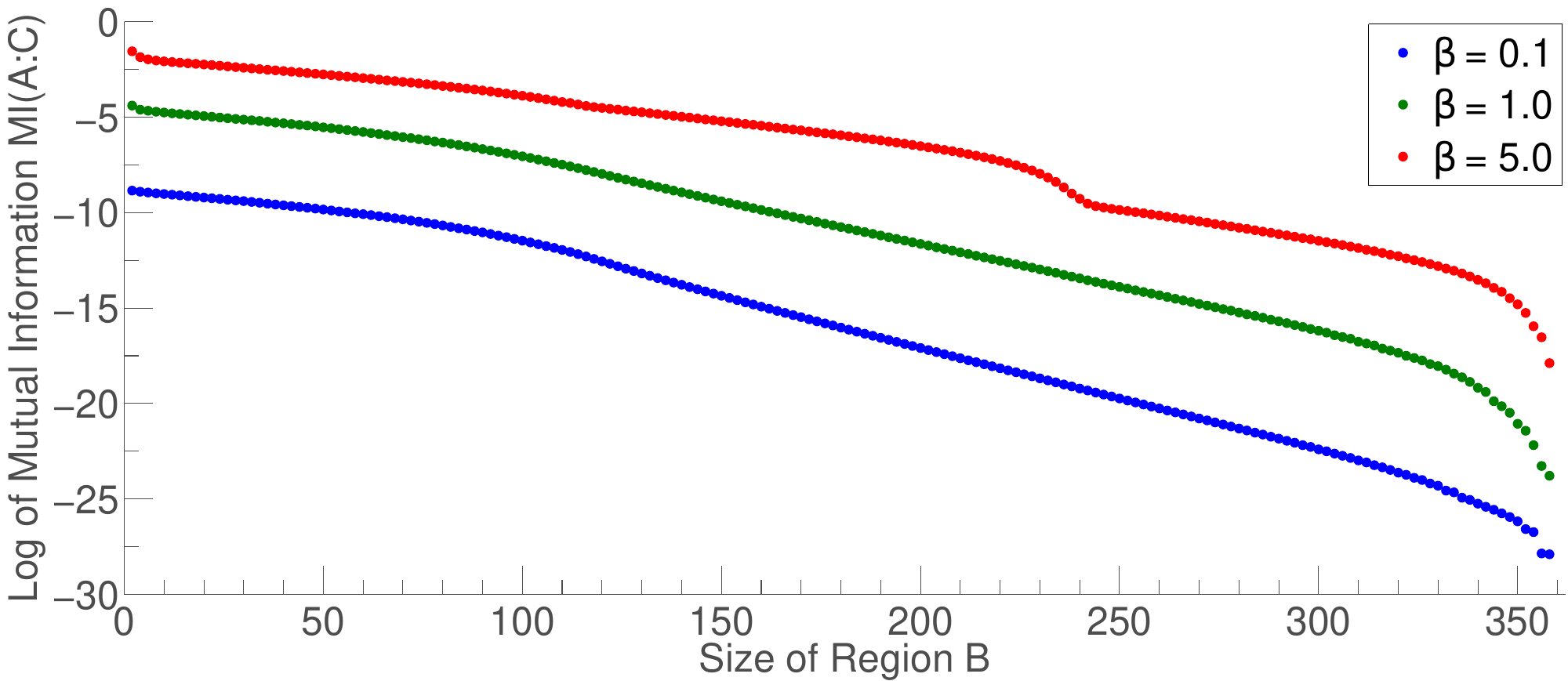}}
	
	\caption{Log of $MI(A:C)$ as a function of separation between regions $A$ and $C$ for different inverse temperatures $\beta$ (a) without ($V_0=0$) and (b) with ($V_0=0.1w$) disorder. Averaging is performed over $500$ disorder realizations. The region geometries are defined in Figure \ref{fig:setup}. Similar results were obtained for the log of $CMI(A:C|B)$.}
	\label{fig:mutual_info}
\end{figure}

We continue our analysis of the wire model by studying the entanglement structure of the steady state. The region configurations we study were defined in Ref.~\cite{mahajan2016entanglement} and are motivated by the physics of approximate condition independence~\cite{2016arXiv160705753S,petz1986,qmarkov,fr1,fr2,2015RSPSA.47150338W,2015arXiv150907127J}. Consider a system with $\mu_R = -\mu_L = 0.1$, sufficiently large leads $N_L=N_R=N_W=120$, and sufficiently small $\gamma = 0.05$ (these choices are based on the parameters that give the best thermalization results). The mutual information $MI(A:C)$ for the partition in Fig.~\ref{fig:setup}, as a function of the size of the middle region $B$ is presented in Fig.~\ref{fig:mutual_info}(a). Different curves correspond to different values of $\beta = \beta_L = \beta_R$. A similar computation for the conditional mutual information shows that it is basically equal to the mutual information, and both decrease exponentially fast. When the logarithm of the MI is less than $-30$, this should be interpreted as the system having $0$ MI. Those quantities are not exactly $0$ because of the numerical precision of our computation. 

\subsubsection{Effects of disorder}

As we briefly mentioned in Sec.~\ref{sec:methods}, we can add a disorder term $V_x c_x^\dagger c_x$ at every site $x$ in the region W of the wire. We choose $V_x$ from a continuous uniform distribution over the interval $[-V_0, V_0]$. We look at small disorder values $V_0 = 0.1w$, although the results of this section hold true for larger $V_0$ as well. For the same system as above with $\mu_R = -\mu_L = 0.1$, $N_L=N_R=N_W=120$, and $\gamma = 0.05$, we plot the mutual information (averaged over $500$ disorder realizations) in Fig.~\ref{fig:mutual_info}(b). A similar computation for the conditional mutual information yields the same result. We see that these quantities decay roughly the same way as in the absence of disorder. 

Similarly, we study the effect of disorder on relaxation rates. For the same system parameters mentioned above, we compute the relaxation rates for $\beta = 1$ and three disorder strengths, $V_0 = 0.1w, 0.5w, w$, and average the results over $1000$ disorder realizations. The results are shown in Fig.~\ref{fig:relaxation_rates}(b). Even though we present the plots only for $\beta = 1$, a similar behavior is observed at both high ($\beta = 0.1$) and low temperatures ($\beta = 5$). We see in this plot the physics of localization taking hold as the time to relax to the steady state now grows exponentially with the total system size.

\subsection{Chern insulator model}
\label{sec:chern}

In this subsection we describe our analysis of the steady state physics of the Chern insulator model. Throughout this discussion we set the nearest neighbor Hamiltonian couplings to be $w=1$ and $w'=0.1$. Other parameters of the model are set to be $V = 3$, $c = 1$, $e_s = 0.5$, $t = 1$, and $\gamma = 0.05$, so that the system has an energy gap with two edge states.
We first study the physics of thermalization when the leads are unbiased. Then we consider the physics of the NESS when the leads are biased. Finally we discuss the structure of entanglement in the steady state. A related study of a bosonic symmetry protected state subject to open system dynamics may be found in Ref.~\cite{2016arXiv160607651R}.

\subsubsection{Unbiased thermal equilibrium}

\begin{figure}
    \centering
    \subfloat[]{\includegraphics[width=\columnwidth]{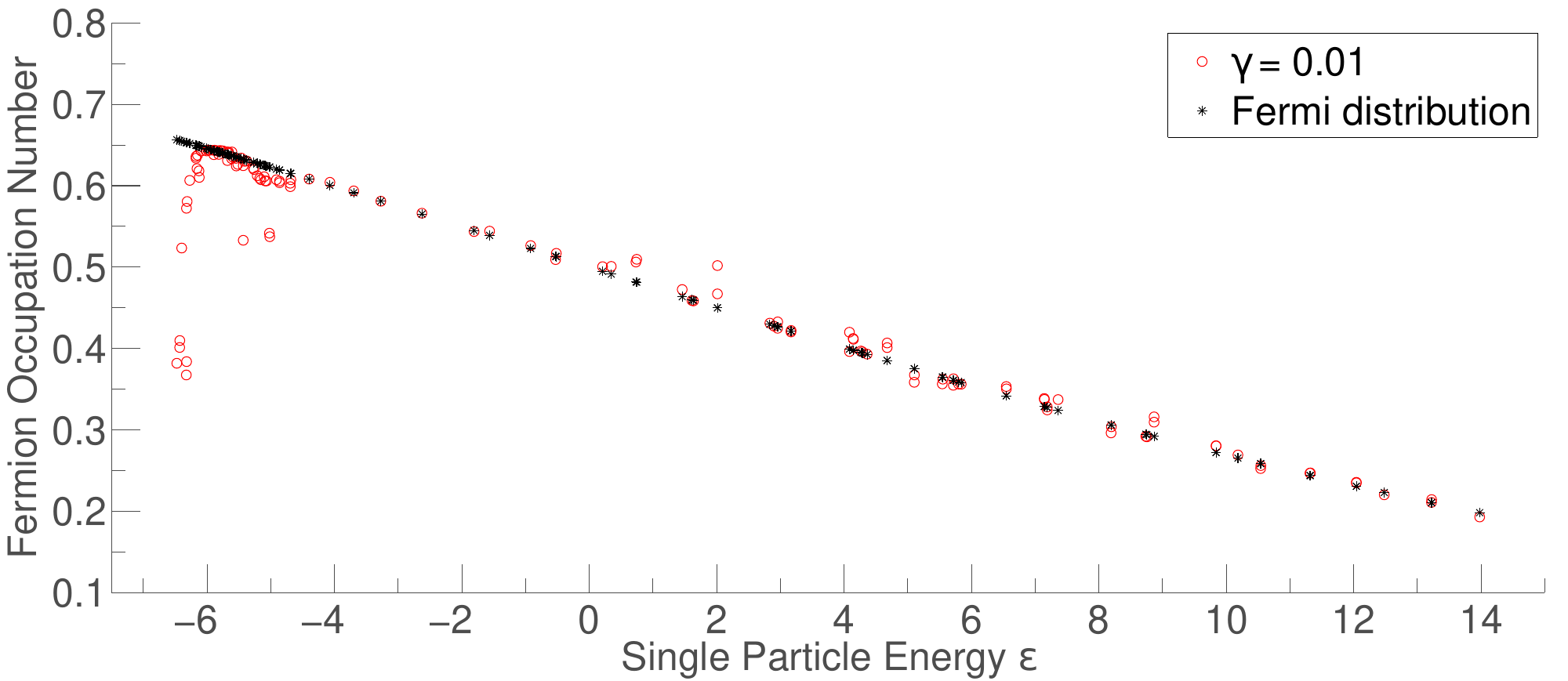}}
    
    \subfloat[]{\includegraphics[width=\columnwidth]{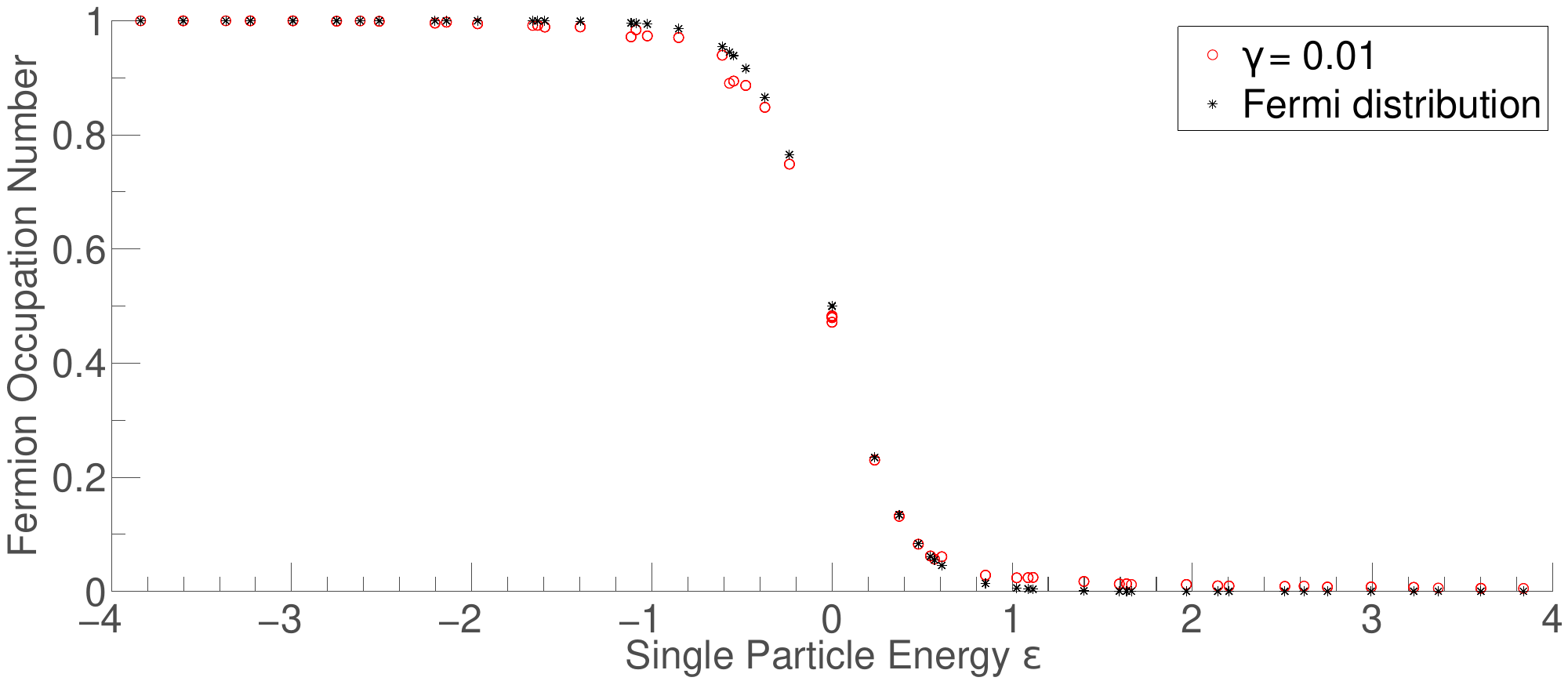}}

	\caption{Comparison of energy eigenstate occupation numbers between thermal equilibrium and the steady state of $\hat{\mathcal{L}}$. Occupation numbers for
	(a) $\beta = 0.1$ and $V = 3$, (b) $\beta = 5$ and $V\rightarrow0$. System sizes are $N_{L, x} = N_{L, y} = N_{W, x} = N_{W, y} = N_{R, x} = N_{R, y} = 10$.}
	\label{fig:occupation_num_2D}
\end{figure}

Here we follow closely our study of the one-dimensional quantum wire, but for the case of the two-dimensional Chern insulator. Consider the unbiased case where $\beta_L = \beta_R = \beta$ and the chemical potential is inside the gap $\mu_L = \mu_R = -3.6$. We find that the leads are able to drive the system to thermal equilibrium in the high temperature limit. The energy occupation numbers for a system with $N_{L, x} = N_{L, y} = N_{W, x} = N_{W, y} = N_{R, x} = N_{R, y} = 10$ and $\beta = 0.1$ are shown in Fig.~\ref{fig:occupation_num_2D}(a).

In the limit $V\rightarrow 0$, the energy gap closes, and the system behaves like an ordinary metal. In this regime, we find that sufficiently large leads can thermalize the system even at low temperatures. The energy occupation numbers for $\beta = 5$ are shown in Fig.~\ref{fig:occupation_num_2D}(b). However, we have not been able to find a parameter regime which thermalizes the Chern insulator with $V\neq 0$ at the lowest temperatures. As we show below, the rate of decay to the steady state is also slow for the Chern insulator, and these observations may be related. 

\subsubsection{Biased steady state}

\begin{figure}
    \centering
    \subfloat[]{\includegraphics[width=\columnwidth]{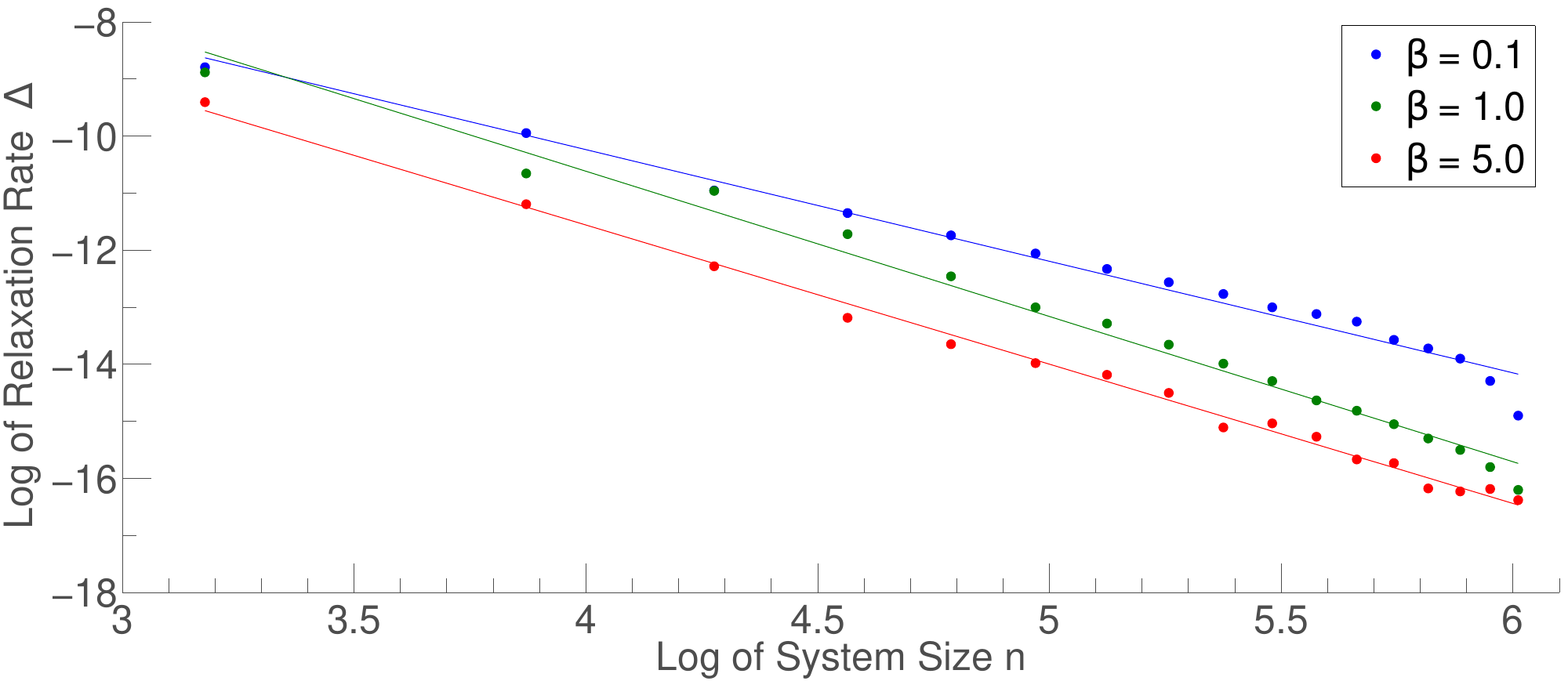}}
    
    \subfloat[]{\includegraphics[width=\columnwidth]{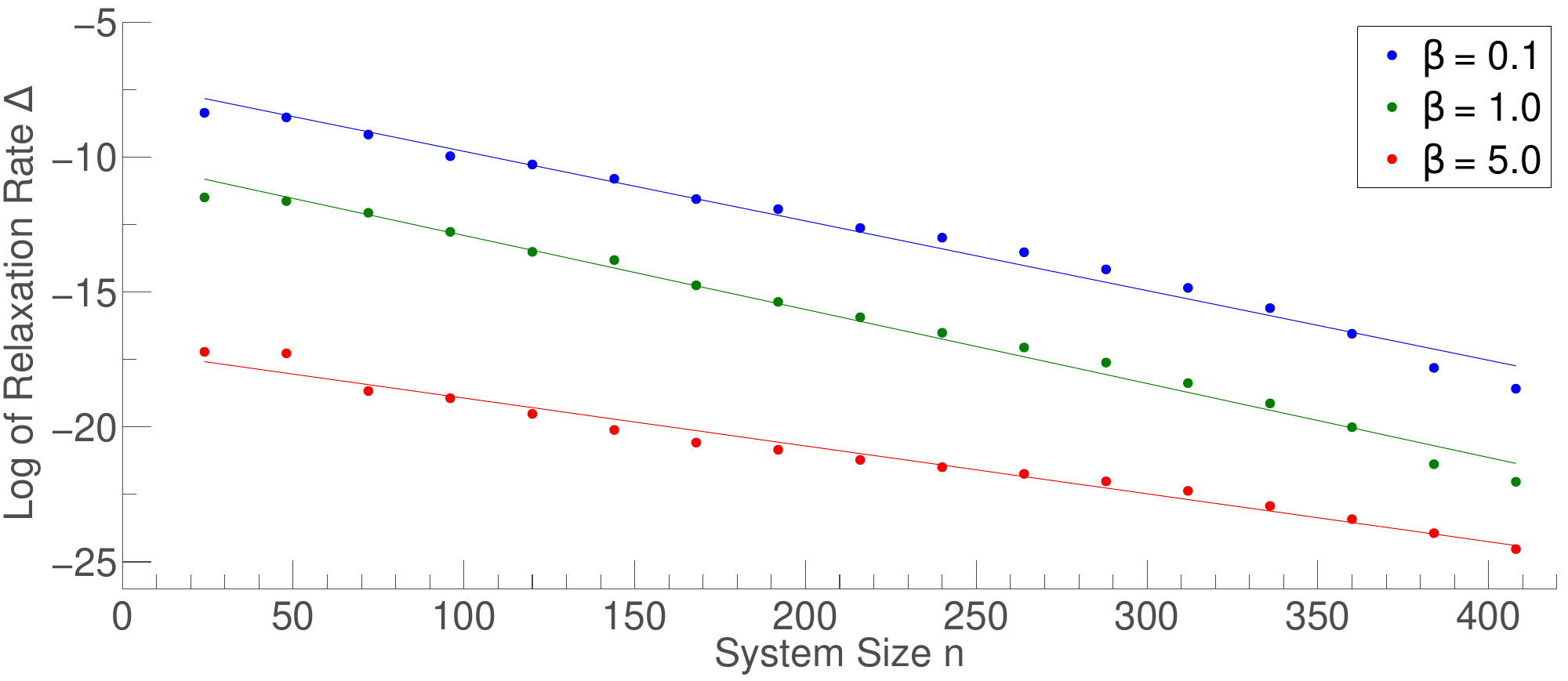}}
	
	\caption{ Log of the relaxation rates $\Delta$ as a function of system size $n=N_{L, x} \cdot N_{L, y} + N_{W, x} \cdot N_{W, y} + N_{R, x} \cdot N_{R, y}$. Relaxation rate for (a) the metallic state and (b) the insulator state and different inverse temperatures $\beta$.
	Note that unlike the metallic state, where $\Delta \sim n^{-a}$, in the insulator state the decay rate $\Delta$ scales exponentially $\Delta \sim e^{-b n}$ with the total system size $n$.}
	\label{fig:relaxation_rates_2D}
\end{figure}

The structure of the current carrying state is generally more complex in the Chern insulator model relative to the quantum wire model because there exist many more current carrying modes in two dimensions. However, when the chemical potential is chosen to sit within the bulk energy gap, then the edge states of the Chern insulator are the primary carriers of current at low temperature. We numerically verify that the current indeed flows primarily on the edges of the bulk. Therefore, we conclude that in addition to driving the system to thermal equilibrium, the leads can also drive the system into a NESS with the expected properties.

We also study the time needed to reach the NESS in the Chern insulator model. We set $N_{L, y} = N_{W, y} = N_{R, y} = 8$ and vary the length of the insulator and leads $N_{L, x} = N_{W, x} = N_{R, x}$. We also fix the chemical potential inside the gap $\mu_L = -3.7$, $\mu_R = -3.5$. Figure~\ref{fig:relaxation_rates_2D} shows the relaxation rate $\Delta$ as a function of total system size for different values of inverse temperature $\beta$. Panel (a) shows that the decay rate for the metallic phase ($V = 0$) depends approximately inverse polynomially on system size. This is the same qualitative structure as obtained from the one-dimensional clean metallic wire. Panel (b) shows that the decay rate depends exponentially on system size in the Chern insulating phase ($V = 3$). This indicates a rather slow approach to the NESS and is similar to the decay rate obtained for the one-dimensional disordered insulator. 

\subsubsection{Entanglement structure of the steady state}

\begin{figure}
    \centering
	\includegraphics[width=\columnwidth]{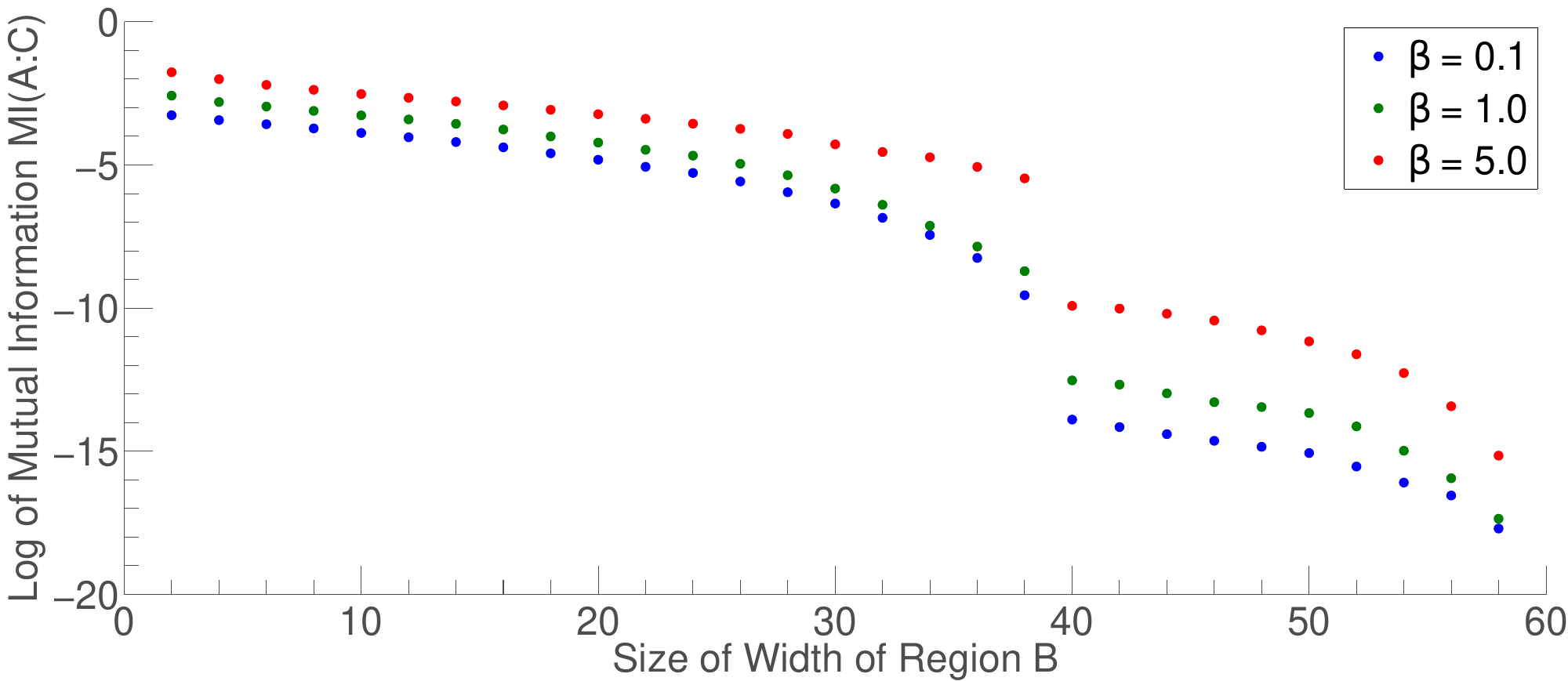}
	\caption{Log of $MI(A:C)$ as a function of x-coordinate separation between regions $A$ and $C$ for different inverse temperatures $\beta$. A similar plot is obtained for the log of $CMI(A:C|B)$.}
	\label{fig:mutual_info_2D}
\end{figure}

Finally, we turn again to an analysis of the entanglement structure of the NESS. Consider a 2D system with $N_{L, x} = N_{R, x} = 10$, $N_{W, x} = 40$, and $N_{L, y} = N_{W, y} = N_{R, y} = 8$. We place the chemical potential in the energy gap $\mu_L = -3.6$, $\mu_R = -3.5$. The mutual information $MI(A:C)$ as a function of the width of the middle region $B$ is plotted in Fig.~\ref{fig:mutual_info_2D}. Note that the discontinuity in the graph occurs exactly when region $B$ becomes the whole insulator, and regions $A$ and $C$ become the left and right lead respectively. As in all other cases, the CMI is basically equal to the MI and both decay exponentially with the size of $B$. Hence the physics of approximate conditional independence survives in the NESS of a two-dimensional model.

\section{Discussion}
\label{sec:discussion}

In this paper we have given detailed evidence that one can design a relatively simple lead that thermalizes target non-interacting fermion systems in one and two dimensions. By biasing the leads, the systems can also be driven out of equilibirum to give expected results. Furthermore, we found that not every lead is able to thermalize a given target, at least in the non-interacting limit considered here. In particular, low temperatures seem to require a larger lead. It is reasonable to conjecture that adding interactions to the target system will help render thermalization more universal, e.g. independent of the details of the bath, but much remains to be understood in this context. In particular, it would be desirable to have a better understanding of what parameters, e.g. lead size and $\gamma$, lead to the best thermalization. More generally, this is just one small step towards a general framework for constructing leads that are able to effectively thermalize a variety of models.

We also showed that the time to reach a steady state is reasonable and scales with an inverse power of the system-plus-lead size, at least for the one-dimensional quantum wire and two-dimensional Chern insulator in its metallic phase. We did find that the environment-lead interaction rate $\gamma$ needs to be small to achieve thermalization. Such a small $\gamma$ further slows the rate of approach to the steady state, but we did not find evidence that $\gamma$ had to scale with system size. In the presence of disorder in one dimension or in the insulating phase in two dimensions, we found that the time to reach a steady state was longer, scaling exponentially with the system size. This presents a potential challenge to the open system dynamics method for finding NESS, although at least in one dimension we know this slowdown is physical and is due to the physics of localization~\cite{PhysRev.109.1492}. Given that larger baths better approximate thermal equilibrium but have slower relaxation times, it would be interesting to investigate optimizing the bath size with respect to these two competing demands. It would also be desirable to investigate the effects of interactions on the relaxation times of the open system dynamics.

Finally, we showed that the systems we investigated have little entanglement in their steady states. Although we used free fermion technology in our calculations, the low degree of entanglement implies that the states in question could also have been represented in a tensor network form. Such a low entanglement structure has been argued to generalize beyond the non-interacting limit~\cite{Prosen2009,mahajan2016entanglement}. It might be illuminating to go through the exercise of writing the non-interacting fermion steady states using tensor networks, particularly as a first step towards including the effects of interactions.

In summary, our results provide further support for the previously outlined entanglement-based approach to calculating electrical and thermal currents in strongly interacting systems. The main idea is to represent the state of the system using tensor network methods and then to dynamically evolve, using an appropriate open system dynamics, to a current-carrying steady state. Here we have shown how to design a non-interacting fermion lead which has three key properties: (1) it effectively thermalizes the system down to low temperatures, (2) it does so in a reasonable time frame, and (3) it is not overly complex. Pioneering matrix product based computations have already been carried out in one dimension at high temperature~\cite{Prosen2009}, and while there are still algorithmic and conceptual barriers to implementing the general program outlined in the introduction, our results have shed light on the design of lead systems. We have also raised several issues in the physics of thermalization of open non-interacting fermion models. Our future work will be concerned with testing these ideas for interacting systems at low temperatures in a variety of dimensions.

\bigskip

\begin{acknowledgments}
CZ is supported by a Stanford UAR Major Grant and by the Clifford C. F. Wong Undergraduate Scholarship Fund. BS is supported by the Simons Foundation as part of the It From Qubit Collaboration. We thank D. Freeman, R. Mahajan, N. Tubman, and P. Hayden for discussions on related issues. We also thank J. Andress for feedback on the manuscript. 
\end{acknowledgments}

\bibliographystyle{apsrev4-2}
\bibliography{references}

\appendix
\section{Solution of the Lindblad equation for quadratic open fermion systems} 
\label{sec:Appendix A}

In this appendix we discuss in more details the method for solving the Lindblad master equation for a system with Hamiltonian and jump operators given by Eqs.~\eqref{eq:Hamiltonian} and \eqref{eq:bath} respectively. First, consider a $4^n$ dimensional Liouville space of operators $\mathcal{K}$, with an inner product defined as follows

\begin{equation}
\langle x|y \rangle = 2^{-n}\mbox{tr}(x^\dagger y), \hphantom{--} x, y \in \mathcal{K}.
\end{equation}
A complete orthonormal basis for this space is given by operator-products

\begin{equation}
P_{\alpha_1, \alpha_2,\ldots, \alpha_{2n}} = w_1^{\alpha_1}w_2^{\alpha_2}\cdots w_{2n}^{\alpha_{2n}}, \hphantom{--} \alpha_j \in \{0, 1\}.
\end{equation}
It turns out that $|P_{\vec{\alpha}}\rangle$ is a fermionic Fock basis, and we can define creation and annihilation linear maps (also known as adjoint Fermi maps) over $\mathcal{K}$

\begin{align}
\hat{c}_j|P_{\vec{\alpha}}\rangle = \alpha_j |w_jP_{\vec{\alpha}}\rangle, && \hat{c}_j^\dagger|P_{\vec{\alpha}}\rangle = (1-\alpha_j) |w_jP_{\vec{\alpha}}\rangle,
\end{align}	
which satisfy the canonical anti-commutation relations

\begin{align}
\{\hat{c}_j, \hat{c}_k\} = 0, && \{\hat{c}_j, \hat{c}_k^\dagger\} = \delta_{j, k}.
\end{align}	
In terms of these maps, the Liouvillean can be written as a quadratic form~\cite{prosen2008third}

\begin{equation}
\label{eq:Liouvillean}
\hat{\mathcal{L}} =  -4i\sum_{j, k=1}^{2n}\hat{c}_j^\dagger H_{jk} \hat{c}_k + \dfrac{1}{2} \sum_{i}\sum_{j, k=1}^{2n}l_{i, j}l_{i, k}^*\hat{\mathcal{L}}_{j, k} ,
\end{equation}
where 

\begin{align}
\begin{split}
\hat{\mathcal{L}}_{j, k} &= (\mathbb{1} + (-1)^{\hat{\mathcal{N}}})(2\hat{c}_j^\dagger\hat{c}_k^\dagger - \hat{c}_j^\dagger\hat{c}_k - \hat{c}_k^\dagger\hat{c}_j)\\
& + (\mathbb{1} - (-1)^{\hat{\mathcal{N}}})(2\hat{c}_j\hat{c}_k - \hat{c}_j\hat{c}_k^\dagger - \hat{c}_k\hat{c}_j^\dagger)
\end{split}
\end{align}
and $\mathcal{N} = \sum_j \hat{c}_j^\dagger \hat{c}_j$ is the number of adjoint fermions. Notice that the Liouvillean commutes with the parity operator $\hat{\mathcal{P}} = (-1)^{\hat{\mathcal{N}}}$ and hence the operator space can be decomposed into even an odd subspaces via an orthogonal projection $\mathcal{K}^\pm = \frac{1}{2} (\mathbb{1} \pm \hat{\mathcal{P}})\mathcal{K} $. Since we are interested in expectation values of observables quadratic in fermion operators, we restrict ourselves to the subspace $\mathcal{K}^+$ where $\hat{\mathcal{L}}_{j, k}$ has the form

\begin{equation}
\label{eq:proj Liouvillean}
\hat{\mathcal{L}}_{j, k|\mathcal{K}^+} = 4\hat{c}_j^\dagger\hat{c}_k^\dagger - 2\hat{c}_j^\dagger\hat{c}_k - 2\hat{c}_k^\dagger\hat{c}_j.
\end{equation}

Combining Eqs.~\eqref{eq:Liouvillean} and~\eqref{eq:proj Liouvillean} yields a compact representation of the Liouvillean on the even subspace

\begin{equation}
\label{eq:pos Liouvillean}
\hat{\mathcal{L}}_{+} = -2\hat{c}^\dagger(2iH+M+M^T)\hat{c} + 2\hat{c}^\dagger(M-M^T)\hat{c}^\dagger ,
\end{equation}
where $\hat{c} = [\hat{c}_1, \hat{c}_2,\ldots, \hat{c}_{2n}]^T$ is a column vector, $M$ is the matrix containing information about the leads defined in Eq.~\eqref{eq:matrix M}, and $H$ is the Hamiltonian matrix with entries given by Eq.~\eqref{eq:Hamiltonian}. We can simplify this representation by introducing $4n$ Hermitian Majorana maps $\hat{a}_k$

\begin{align}
\hat{a}_{2j-1}=\dfrac{1}{\sqrt{2}} (\hat{c}_j+\hat{c}_j^\dagger), && \hat{a}_{2j}=\dfrac{i}{\sqrt{2}} (\hat{c}_j-\hat{c}_j^\dagger).
\end{align}	
In terms of these maps, Eq.~\eqref{eq:pos Liouvillean} becomes 

\begin{equation}
\hat{\mathcal{L}}_{+} = \hat{a}A\hat{a} - A_0\mathbb{1} ,
\end{equation}
where $\hat{a} = [\hat{a}_1, \hat{a}_2,\ldots, \hat{a}_{4n}]^T$ is a column vector, $A$ is the antisymmetric matrix introduced in Eq.~\eqref{eq:matrix A}, and $A_0 = 2\mbox{tr}(M)$. 

Assuming that $A$ is diagonalizable with eigenvalues $\beta_1, -\beta_1, \beta_2, -\beta_2, \ldots, \beta_{2n}, -\beta_{2n}$ and eigenvectors $v_1, v_2, \ldots, v_{4n}$, the Liouvillean can be written in normal form

\begin{equation}
\hat{\mathcal{L}}_{+} = -2\sum_{j=1}^{2n}\beta_j\hat{b}_j'\hat{b}_j ,
\end{equation}
where $\hat{b}_j$ and $\hat{b}_j'$ are normal master mode maps defined as 

\begin{align}
\hat{b}_j = v_{2j-1}^T\hat{a}, && \hat{b}_j'=v_{2j}^T \hat{a}.
\end{align}	

This normal form representation of the Liouvillean leads to the results stated in Section~\ref{sec:methods}. A more detailed derivation of the formulas above is presented in Ref.~\cite{prosen2008third}.

\end{document}